\documentclass[10pt,emulateapj,apj]{emulateapj}
\newcommand{\fe}{$f_E $}

\shorttitle{Galaxy Interactions and Large-Scale Environment}
\shortauthors{Park \& Choi}
\begin{document}
\title{Combined Effects of Galaxy Interactions and \\
Large-Scale Environment on Galaxy Properties}
%\twocolumn[
\author{ Changbom Park\altaffilmark{1}, 
\& Yun-Young Choi\altaffilmark{2}}
\altaffiltext{1}{Korea Institute for Advanced Study, Dongdaemun-gu, Seoul 130-722, Korea; cbp@kias.re.kr}
\altaffiltext{2}{Astrophysical Research Center for the Structure and
Evolution of the Cosmos, Sejong University, 
Seoul 143-747, Korea; yychoi@kias.re.kr}

\begin{abstract}
We inspect the coupled dependence of physical parameters of the Sloan 
Digital Sky Survey galaxies
on the small-scale (distance to and morphology of the nearest
neighbor galaxy) and the large-scale (background density smoothed over
20 nearby galaxies) environments.
The impacts of interaction on galaxy properties are detected at least
out to the neighbor separation corresponding to the virial radius of galaxies, 
which is typically between 200 and 400 $h^{-1}$ kpc for the galaxies in our sample.
To detect these long-range interaction effects
it is crucial to divide galaxy interactions into four cases dividing the
morphology of target and neighbor galaxies into early and late types.
We show that there are two characteristic neighbor-separation scales where 
the galaxy interactions cause abrupt changes in the properties of galaxies.
The first scale is the virial radius of the nearest neighbor galaxy $r_{\rm vir,nei}$.
Many physical parameters
start to deviate from those of extremely isolated galaxies
at the projected neighbor separation $r_p$ of about $r_{\rm vir,nei}$.
The second scale is at $r_p \approx 0.05 r_{\rm vir,nei} = 10 -20 h^{-1}$ kpc,
and is the scale at which the galaxies in pairs start to merge.
We find that late-type neighbors enhance the star formation activity of galaxies 
while early-type neighbors reduce it, and 
that these effects occur within $r_{\rm vir,nei}$.
The hot halo gas and cold disk gas must be participating in the interactions
at separations less than the virial radius of the galaxy plus dark halo system.
Our results also show that the role of the large-scale density in determining 
galaxy properties is minimal once luminosity and morphology are fixed. 
We propose that the weak residual dependence
of galaxy properties on the large-scale density is due to
the dependence of the halo gas property on the large-scale density.

\end{abstract}

\keywords{galaxies:general -- galaxies:formation
-- galaxies:evolution -- galaxies:morphology -- galaxies:properties}

\section{Introduction}
According to the currently popular $\Lambda$CDM model of structure formation, 
galaxies should form hierarchically. 
Numerical simulations demonstrate that galaxy-sized 
objects form through numerous interactions and mergers (e.g. Toomre \&
Toomre 1972; Hernquist 1992, 1993; Naab \& Burkert 2003; Robertson et al. 2006; Maller et al. 2006). 
Therefore, the outcomes of galaxy-galaxy interactions 
and mergers are of key importance in understanding the hierarchical picture
of galaxy formation.
However, the impact of interactions and 
mergers on galaxy properties is still not known well even though 
there have been many studies.

One of the earliest works on this problem is
Larson \& Tinsley (1978) who showed that galaxies selected from the Atlas of
Peculiar Galaxies (Arp 1966) have enhanced star formation (SF) rate
compared to typical galaxies.
Since that time, many studies have shown 
observational evidence of significant changes in galaxy properties
due to interactions and/or mergers, for instance, 
in SF activity (Kennicutt \& Keel 1984; 
Kennicutt et al. 1987; Bushouse, Werner \& Lamb 1988;
Barton et al. 2000, 2003, 2007; Lambas et al. 2003; 
Sanchez \& Gonzalez-Serrano 2003; 
Nikolic, Cullen \& Alexander 2004; Hernandez-Toledo et al. 2005;
Geller et al. 2006; Li et al. 2008),
%Kennicutt et al. (1987) reported a general trend for enhanced star formation 
%and nuclear activity of interacting galaxies although with a wide dispersion.
%Bushouse, Werner \& Lamb (1988) reported enhanced emission of the FIR in some, 
%but not all, interacting pairs.
galaxy structure parameters (Nikolic et al. 2004;
Patton et al. 2005; Hernandez-Toledo et al. 2005; 
Coziol \& Plauchu-Frayn 2007; Kacprzak et al. 2007),
luminosity ratio (Woods et al. 2006; Woods \& Geller 2007),
and stellar mass ratio (Ellison et al. 2008). 
Recent  $N$-body simulations have also shown that 
the general features of the observations can be modeled 
(e.g. Barnes \& Hernquist 1996; Mihos \& Hernquist 1996; Tissera et al. 2002; 
Perez et al. 2006a,b; Di Matteo et al. 2007).

However, previous studies have not always been in agreement with one
another. 
Yee \& Ellingson (1995) and Patton et al. (1997) found no significant 
difference between the mean properties of isolated and paired galaxies.
Bergvall et al. (2003) found no clear difference between isolated 
galaxies and interacting/merging systems in their global SF rate in 
Optical/near-IR bands.
Allam et al. (2004) identified a set of merging galaxies in the SDSS data
and found only a weak positive correlation in color for the merging pairs.
Hernandez-Toledo et al. (2005) analyzed the $BVRI$ images of 42 
elliptical/lenticular galaxies, 
and claimed that the structural effects of interactions on E/S0s 
are minor, in contrast to disk galaxies involved in interactions.
de Propris et al. (2005) analyzed galaxies in the Millennium Galaxy 
Catalog and found merging galaxies are only marginally bluer 
than noninteracting galaxies, showing an excess of both early 
and late types but a deficiency of intermediate type spirals.
Smith et al. (2007) did not find any 
enhancement in Spitzer mid-infrared color depending on pair separation  
by using pre-merger interacting galaxy pairs selected from the Arp Atlas.

In addition, 
in terms of the degree and scale of the effects of interactions on
galaxy properties, 
the results also have not always agreed.
Lambas et al. (2003) studied the galaxy pairs in the Two Degree Field 
Galaxy Redshift Survey (2dFGRS) data and reported that SF
in galaxy pairs is significantly enhanced over that of isolated galaxies 
only when the projected
separation $r_p<25 h^{-1}$ kpc and radial velocity difference 
$\Delta v<100$ km s$^{-1}$.
Nikolic et al. (2004) reported using the SDSS data that
the mean SF rate is significantly enhanced for 
galaxy pairs with $r_p < 30$ kpc. 
But for late types, the enhancement extended out to 300 kpc.
They also noted that the SF rate slightly decreased with increasing $\Delta v$, and
the light concentration was lowest at $r_p=75$kpc and then increased rapidly inward. 
Alonso et al. (2006) measured SF rate of the galaxies in the 2dFGRS and SDSS data
and found that the SF rate was strongly enhanced when $r_p <100 h^{-1}$ kpc and 
$\Delta V<350$ km s$^{-1}$, which was more effective in low and intermediate
density environments.
These discrepancies as listed above are expected mainly because the effects of 
interactions/mergers between different types of galaxies on their 
final products are different. 
Woods \& Geller (2007) found that for blue-blue major pair sample 
in the SDSS, there exists clear correlation between specific SFR and pair 
separation, and for red-red pair, there is none (cf. Tran et al. 2005;
van Dokkum 2005; Bell et al. 2006).
Li et al. (2008) found that for the most strongly star-forming systems,
tidal interactions are the dominant trigger of enhanced
star formation and the enhancement is a strong function of separation
less than 100 kpc. 

In addition to such small-scale environment, the large-scale environment
has been known to be one of the determining factors of galaxy properties.
It is now well-known that galaxy properties 
correlate with the large-scale background density at low redshift
(Hogg et al. 2003; Goto et al. 2003; Balogh et al. 2004a,b;
Tanaka et al. 2004; Kauffmann et al. 2004; 
Blanton et al. 2005a; Croton et al. 2005; 
Weinmann et al. 2006; Park et al. 2007). Deep redshift surveys 
have extended these studies to high redshift (Cucciati et al. 2006; 
Elbaz et al. 2007; Cooper et al. 2007, 2008; Poggianti et al. 2008).
%
%Hogg et al. (2004) found that the color-magnitude relation of bulge-dominated
%galaxies is independent of the large-scale density.
%
A number of papers (Kauffmann et al. 2004; Blanton et al. 2005a;
Quintero et al. 2006; Ball, Loveday, \& Brunner 2008) 
claimed from an analysis of the SDSS
data that structural properties of galaxies are less closely 
related to the (large-scale) environment than are their masses and SF 
related parameters such as color and SF rate.
However, the structural parameters such as concentration and Sersic index
are not true measure of morphology (Bamford et al. 2008) and
indeed, van der Wel (2008) pointed
that morphology and structure are intrinsically different galaxy properties
and structure mainly depends on galaxy mass whereas morphology mainly
depends on environment. 

Several papers tried to use sophisticated
automated morphology classifications for the study of the relationship between
morphology and environment 
(Goto et al. 2003; Park \& Choi 2005; Allen et al. 2006; Ball et al. 2008).
% and that color is a more 
%fundamental predictor of environment than morphology.
%Baldry et al. (2006) found that the fraction of red galaxies
%increases as the projected neighbor density and the stellar mass increase,
%and that the loci of red and blue sequences in color-mass and
%color-concentration index are insensitive to environment,
%which has also been found by Balogh et al. (2004a).
Many studies reported that the SF rate of galaxies is
a strongly decreasing function of the large-scale density and that there is
a critical density for SF activity (Gomez et al. 2003; Tanaka 
et al. 2004).

Park et al. (2007), however, found that this trend was mostly because morphology 
and luminosity are strong functions of the large-scale density. 
They made an extensive study of the environmental dependence
of various physical properties of galaxies on large- and small-scale 
densities, and concluded 
that morphology and luminosity are major fundamental parameters.
Once morphology and luminosity are fixed, 
other galaxy properties are almost independent of the large-scale density.
The large-scale density dependence of these properties reported
by many previous and current studies merely reflects the correlations
of the properties with morphology and luminosity rather than independent 
correlations with environment.
Park et al. (2007) also found that galaxy morphology changes sensitively
across the nearest neighbor distance of a few hundred kpc.
This work has been extended by Park, Gott, \& Choi (2008) who studied 
galaxy morphology as a function of the nearest neighbor separation
at fixed large-scale background density and found
that this characteristic scale corresponds to the virial radius of
the neighbor galaxy. 
Park et al. (2008)'s findings and claims can be summarized as follows.
%\begin{description}
%\item 

1. The effects of galaxy interactions reach farther than the
distance previously thought. 
They reach at least out to the galaxy virial radius, namely 
a few hundred kpc for bright galaxies.

%\item 
2. The result of galaxy interactions can be very different depending on 
the morphological type of the nearest neighbor galaxy when the pair separation is 
less than the virial radius. 
Without separating the neighbor galaxies into different morphological types, 
one will find the effects of galaxy interactions are diverse but negligible 
on average. The dependence on neighbor's morphology disappears 
at separations farther than the virial radius.

%\item 
3. The fact that, at fixed large-scale density,
the morphology of a galaxy is more likely to be a late type 
as it approaches a late-type neighbor, suggests that galaxies can transform 
their morphology from early types to late types through close encounters 
with cold gas-rich neighbors. 

%\item 
4.  
In most places of the universe, except for the regions within massive
clusters of galaxies, the well-known morphology-density relation 
originated largely due to the effects of galaxy-galaxy interactions. 
Galaxy morphology appears to depend on the large-scale density mainly 
because the mean separation between galaxies is statistically correlated 
with the large-scale density. 
%The large-scale density seems to affect 
%the hot halo gas of interacting galaxies and change galaxy morphology 
%only indirectly. Hydrodynamic processes like ram pressure stripping, 
%viscous stripping, and thermal evaporation, may be the physical mechanisms 
%that are responsible for the evolution of galaxies interacting within
%each other's virial radius.

%\item 
5. A series of close interactions and mergers transform galaxy 
morphology and luminosity classes in such a way that galaxies on average 
evolve to become bright early types. The transformation speed depends 
on the large-scale density. 
%\end{description}

The fourth result is supplemented by the recent work of Park \& Hwang (2008) 
who found, in the special case of galaxies located within the virial radius of massive
clusters, galaxy properties depend on both the distance to the nearest neighbor 
and  clustercentric radius.

The purpose of this paper is 
to extend Park et al. (2008)'s work by looking at the dependence 
of various other properties of galaxies, as well as morphology and luminosity,
on small and large scale environmental factors.
We use a volume-limited sample of the SDSS galaxies whose morphology is
accurately classified.
The environment is specified by the small-scale (distance to the nearest 
neighbor galaxy and morphology of the nearest 
neighbor galaxy) and large-scale (the background density) factors.
It is hoped that the effects of galaxy interactions can be understood in fuller
detail in this three-dimensional environmental parameter space.

\section{Observational Data Set}
\subsection{Sloan Digital Sky Survey Sample}

\begin{figure} 
\center
%\plotone{fig1.eps}
\includegraphics[scale=0.45]{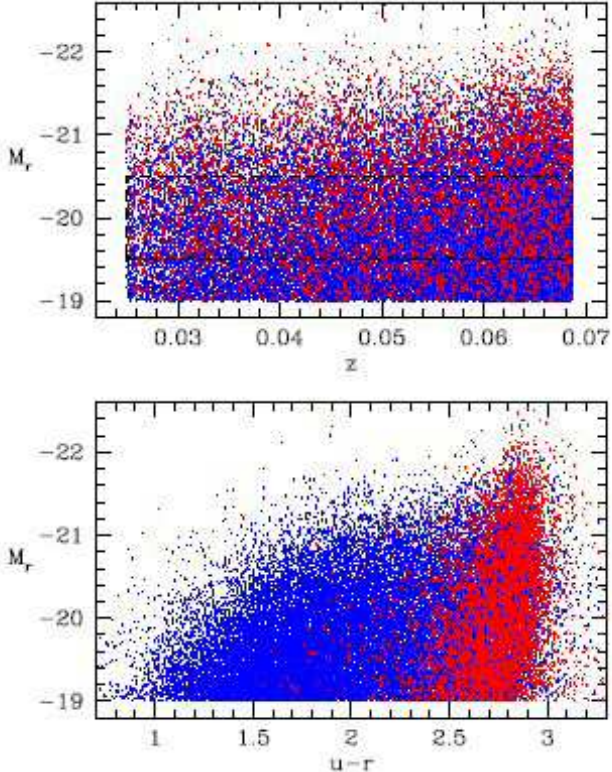}
\caption{The upper panel show all 49,571 galaxies in our volume-limited 
sample, D3.
The rectangular box encloses the target galaxies on which our analysis
is focused. The faint limit of these target galaxies is 0.5 magnitude
brighter than the full sample to achieve complete neighbor 
selection.  The bottom panel shows the galaxies in the color-magnitude diagram.
Red points are early morphological type galaxies, and blue points are
late types.}
\end{figure}

Our observational sample is one of the subsamples, D3, used by Choi et al. 
(2007) and Park et al. (2007). It is a volume-limited sample of
galaxies extracted from a large-scale structure sample, DR4plus (LSS-DR4plus), 
of the SDSS data (York et al. 2000)
from the New York University Value-Added Galaxy Catalog 
(NYU-VAGC; Blanton et al. 2005b). 
This sample is a subset of the spectroscopic Main Galaxy sample of
the SDSS Data Release 5 
(Adelman-McCarthy et al. 2007). 

The sample D3, together with other volume-limited samples, has been
described in great detail by Choi et al. To summarize it is a sample of
galaxies with the $r$-band absolute magnitude $M_r<-19.0+5{\rm log} h$ 
(hereafter we drop the
$+5{\rm log} h$ term in the absolute magnitude) and redshifts $0.025<
z<0.06869$ or comoving distance of $74.6 h^{-1}$Mpc $<d< 203.0 h^{-1}$Mpc.
The SDSS spectroscopic sample has a bright apparent magnitude limit
of $r=14.5$, but our sample is supplemented by brighter galaxies whose redshifts are
obtained from various literature. D3
includes 49,571 galaxies. The rest-frame absolute magnitudes of
individual galaxies are computed in fixed bandpasses, shifted to $z=0.1$,
using Galactic reddening correction (Schlegel et al. 1998) and $K$-corrections
as described by Blanton et al. (2003). The mean evolution correction given by 
Tegmark et al. (2004), $E(z) = 1.6(z-0.1)$, is also applied. 
We adopt a flat $\Lambda$CDM cosmology with
$\Omega_{\Lambda}=0.73$ and $\Omega_m=0.27$. The useful survey area of
this sample, having nonzero angular selection function, is 1.362 sr.
All galaxies in D3 are plotted in Figure 1.

\subsection{Morphology Classification}

Accurate morphology classification is critical in this work since the effects
of interaction depend strongly on morphology of the target and neighbor
galaxies.
We first classify morphological types of galaxies using the prescription of
Park \& Choi (2005). Galaxies are divided into early (ellipticals and 
lenticulars) and late (spirals and irregulars) morphological types 
based on their locations
in the $u-r$ color versus $g-i$ color gradient space and also in the 
$i$-band concentration index space. 
The resulting morphological classification has completeness and 
reliability reaching 90\%.

Our automatic classification scheme does not perform well
when an early-type galaxy starts to overlap with other galaxy.
This is because the scheme excludes galaxies with very low concentration
from the early-type class and blended images often erroneously give low
concentration.
Since we are investigating the effects of close interaction on galaxy
properties, this problem in the automatic classification 
has to be remedied.
We perform an additional visual
check of the color images of galaxies to correct misclassifications 
by the automated scheme for about $10,000$ galaxies
having close neighbors.
In this procedure we changed the types of the 
blended or merging galaxies, blue but elliptical-shaped galaxies,  
and dusty edge-on spirals.
Some non-sense objects like blank fields and substructures of large galaxies,
are removed from the samples, and some wrong
central positions of merging galaxies are corrected. 

After all these procedures
our final sample is composed of 19,248 early-type galaxies, and 30,283
late-type galaxies with $M_r <-19.0$. Our main target galaxies for which we
study the dependence of various physical parameters on environment,
are those with $-19.5>M_r>-20.5$. There are 9,434 early types and
14,270 late types satisfying this condition within the sample volume.
This subset is marked by a rectangular box in Figure 1.
In our analysis we often limit the late-type galaxy sample to those with 
isophotal axis ratio $b/a$ greater than 0.6. This is to reduce the effects
of internal extinction on our results. The absolute magnitude and color of
late-type galaxies with $b/a < 0.6$ are very inaccurate (see Fig. 5 and 12 of
Choi et al. 2007), and including them in the analysis introduces 
a large dispersion in luminosity of the volume-limited sample.  Since it is
essential to fix luminosity in many of our analyses, it is very important to
reduce the internal extinction effects by using nearly face-on late-type
galaxies. When we calculate the median values of galaxy parameters
in section 3, we will take into account the fact that only a subset of 
late-type galaxies are being used.   
There are 8,344 late types with $b/a \ge 0.6$ and $-19.5>M_r>-20.5$
in our sample. We also often divide our sample into four subsamples:
early types having  early-type nearest neighbor (the E-e galaxies),
early types having   late-type nearest neighbor (E-l),
late  types having  early-type nearest neighbor (L-e), and
late  types having   late-type nearest neighbor (L-l).
There are 4675, 4759, 3423, and 4921 galaxies in these subsets, 
respectively, when only those with $b/a>0.6$ are counted in the case
of late-type target galaxies.

\subsection{Local environment}

We consider three kinds of environmental factors. One is
the mass density described by many neighboring galaxies over a few Mpc scale.
This is called the large-scale background density.
Another is the small-scale mass density attributed to the closest neighbor
galaxy.
The third is the morphology of the closest neighbor galaxy.

The background density at a given location of a galaxy
is measured by 
\begin{equation}
\rho_{20}({\bf x})/{\bar\rho} = \sum_{i=1}^{20} \gamma_i L_i W_i(|{\bf x}_i -
{\bf x}|)/{\bar\rho},
\end{equation}
using the $r$-band luminosity $L$ of the closest twenty galaxies 
in the sample. Here ${\bar\rho}$ is the mean density of the
universe, $\gamma$ is the mass-to-light ratio of a galaxy, and
$W(x)$ is a smoothing filter function. 
Here the mass associated with a galaxy plus dark halo system is assumed 
to be proportional to the $r$-band luminosity of the galaxy. 
The mean mass density within a sample of the total volume $V$
is obtained by
\begin{equation}
{\bar\rho} = \sum_{\rm all} \gamma_i L_i /V,
\end{equation}
where the summation is over all galaxies brighter than $M_r = -19.0$  
in the sample.
Only the relative mass--to--light ratios $\gamma$ for early and 
late types are needed in equation (1) since $\gamma$'s appear
both in the numerator and denominator.
We assume $\gamma({\rm early}) = 2\gamma({\rm late})$
at the same $r$-band luminosity. This is our choice of the connection
of luminosity and morphology with the halo mass.
It is based on the results given in section 2.5.
We assume $\gamma$ is constant with galaxy luminosity for a given
morphological type. The mass-to-light ratio of galaxies is actually 
expected to be a monotonically increasing function
of galaxy halo mass over the luminosity range of our sample ($M_r < -19.0$;
see Figures 3 and 4 of Kim, Park, \& Choi 2008).
Then the overdensity of the high density regions will be underestimated.
However, the relation between our mass density estimate and the true one
is still monotonic, and only the labels of $\rho_{20}$ and $r_p$ will change.
We find the mean mass density
\begin{equation}
{\bar \rho} = (0.0223 \pm 0.0005) (\gamma L)_{-20},
\end{equation}
where $(\gamma L)_{-20}$
is the mass of a late-type galaxy with $M_r = -20$.

We use the spline-kernel weight $W(r)$ for the background density 
estimation.
We vary the size of the spline kernel to include
twenty galaxies with $M_r<-19.0$ within the kernel weighting.  
The spline kernel is adopted because it is centrally
weighted, unlike the tophat or cylindrical kernel, and 
has a finite tail, unlike the Gaussian. Our kernel is also adaptive.  
An adaptive kernel constrained to include a fixed number of objects, 
allows more uniform smoothing in the
`initial' conditions compared to the method adopting a fixed-scale
at the present epoch since the high density regions collapse while the 
under-dense regions expand as the universe evolves.
The methods of calculating $\rho_{20}$  and the examination
of the results are described in full detail by Park et al. (2008). 
Interested readers should refer to section 2.3 of the paper.

Our background density estimate $\rho_{20}$ spans the large-scale
environment from voids to clusters. But the smoothing scale determined by
20 galaxies with $M_r<-19.0$ is larger than the typical cluster virial radius,
which is 1--2 $h^{-1}$Mpcs. Therefore, in our calculation $\rho_{20}$ 
never exceeds the virialization density $200\rho_c \approx 740{\bar\rho}$, 
where $\rho_c$ is the critical density of the universe. Our sample 
includes massive clusters, and $\rho_{20}/{\bar\rho}$ at the locations of cluster
member galaxies ranges roughly from 50 to 400. 
At these densities cluster member galaxies are mixed with those outside
clusters.
In our forthcoming paper Park \& Hwang (2008)
we will resolve the virialized regions of Abell clusters, and study
the dependence of galaxies within 10 times the cluster virial radius
on the clustercentric radius and the nearest neighbor distance.

Our background density estimation is made using a spherically symmetric
smoothing kernel and using the redshift space distribution of galaxies.
In redshift space
some of cluster member galaxies are stretched along the line
of sight, appearing as fingers-of-god. This causes smearing of cluster galaxies
into low density regions. Even though the fraction of galaxies dislocated by
the redshift-space distortion more than our smoothing scale is small, a caution
must be given to cases showing a very weak dependence on $\rho_{20}$
in our results.

The small-scale density experienced by a target galaxy 
attributed to its neighbor is estimated by
\begin{equation}
\rho_{n}/{\bar\rho} = 3\gamma_n L_n /4\pi r_p^3 {\bar\rho},
\end{equation}
where $r_p$ is the projected separation of the nearest neighbor galaxy 
from the target galaxy. The density due to the nearest neighbor used
in our work does not represent the galaxy number density at small-scales,
but rather the local mass density given by the nearest neighbor itself.
The method to find the nearest neighbor is described
in the next section. 

We will study the dependence of galaxy properties on the nearest neighbor
distance normalized
by the virial radius of the nearest neighbor. We define the virial radius of a 
galaxy as the projected radius where the mean mass density $\rho_n$
within the sphere with radius of $r_p$ is 200 times the critical density 
or 740 times the mean density of the universe, namely,
\begin{equation}
r_{\rm vir} = (3 \gamma L /4\pi {\rho_c}/200)^{1/3}.
\end{equation}
Since we adopt $\Omega_m = 0.27$, 
$200\rho_c = 200 {\bar\rho}/\Omega_m = 740{\bar\rho}$.
This is almost equal to the virialized density 
$\rho_{\rm virial}=18 \pi^2 / \Omega_m (H_0 t_0)^2 {\bar\rho}
= 766{\bar\rho}$   in the case of our $\Lambda$CDM  universe (Gott \& Rees 1975).
This is what Park et al. (2008) used to define the virial radius.
According to our formulae the virial radii of galaxies with $M_r=-19.5,
-20.0,$ and $-20.5$ are 260, 300, and 350 $h^{-1}$ kpc for early types, 
and  210, 240, and 280 $h^{-1}$ kpc for late types, respectively.

These sizes are much larger than the visible part of galaxies. Therefore,
when we mention a galaxy, we actually mean the galaxy plus dark halo system.
In the next section we will see the importance of the galaxy-component
of the galaxy-halo systems during interactions. 
Besides the gravitational effects,
the halo determines the size of the virial radius of a system and the domain
of hydrodynamic influence. The galaxy determines which kind of hydrodynamic
effects to occur.

\subsection{The nearest neighbor}
We define the nearest neighbor galaxy of a target galaxy as the one 
which is located closest to the galaxy on the sky and
satisfies magnitude and radial velocity conditions.
Suppose we are looking for the neighbors of a target galaxy with $M_r$ 
and with a certain morphological type.
Its neighbors are defined as those with the absolute magnitudes
brighter than $M_r+\Delta M_r$ and the radial velocity difference less than
$\Delta v$. We adopt $\Delta M_r =0.5$ and $\Delta v =$600 km s$^{-1}$
 (early-type target) or 400 km s$^{-1}$ (late-type target).

Our results below are insensitive to the choice of $\Delta M_r$. But if
$\Delta M_r$ is too large, the size of target galaxy sample becomes too small
because the absolute magnitude limit of the full sample has the limit of
$M_r <-19.0$ and the target sample must have the limit of $M_r <-19.0-\Delta M_r$
to be complete in neighbor sampling. We choose $\Delta M_r=0.5$ as the optimum case
making the influential neighbors included and yet
yielding good statistics. 
The galaxies within the rectangular box of Figure 1 is our major
target galaxy sample.
The reason that we use different limits to $\Delta v$ for early and late-type
targets is explained in the next section.

Instead of the nearest one we have also tried to use the most-influential neighbor
that produces the highest local density $\rho_n$
at the location of the target galaxy. 
Our results in the following sections are qualitatively the same when we use
the most-influential neighbor instead of the nearest neighbor because the majority of
the most-influential neighbors are actually the nearest ones.

\subsection{Velocity difference between neighboring galaxies}

\begin{figure} 
\center
\plotone{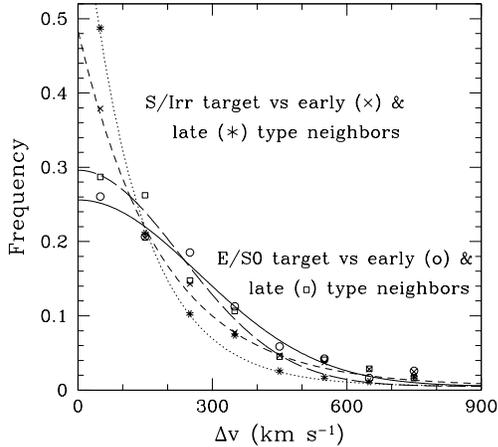}
\caption{Distributions of radial 
velocity difference between the target galaxies with $-19.5>M_r>-20.5$
and their neighbors brighter than $M_r +0.5$ and with separations
$10 h^{-1}{\rm kpc} <r_p < 100 h^{-1}$kpc. 
Cases are distinguished among those of early-type target versus early-type
neighbor ($\circ$) or versus late-type neighbor ($\Box$), and of 
late-type target versus early-type neighbor ($\times$) or versus 
late-type neighbor ($\ast$). Curves are best-fit Gaussian (early-type target cases)
and exponential (late-type target cases) functions.}
\end{figure}

The choice of $\Delta v$ is based on the pairwise
velocity difference between target galaxies and their neighbors 
(see also section 2.4 of Park et al. 2008).
For a given target galaxy with a given morphological type and with $-19.5>M_r>-20.5$ 
we searched for all neighbors with the projected separation of
$10 h^{-1}{\rm kpc} <r_p < 100 h^{-1}$kpc. 
Figure 2 shows the distributions of the
radial velocity difference between the target and neighbor for four cases;
early-type target and early-type neighbor (circles), 
early-type target and late-type neighbor (squares), 
late-type target and early-type neighbor (crosses), 
and late-type target and late-type neighbor  (stars).  
It is important to note that the velocity difference distributions
for late-type target galaxies are quite different from those for
early-type target galaxies. At fixed luminosity late types have smaller
velocity difference with their neighbors than early types.
A sample of galaxy pairs with small $\Delta v$ (say, $<50$ km s$^{-1}$)
will be dominated by late-type galaxies, and that with large $\Delta v$
(say, $>200$ km s$^{-1}$) by early-type galaxies (see the discussion section
for its systematic effects).

As can be seen in Figure 2, the velocity dispersion of neighboring galaxies
around a dark halo is not simply defined since it depends on the morphology
of the neighbors.
The distributions can be fit well by the Gaussian function 
(solid and long-dashed curves) for early-type target galaxies, but by the
exponential function (short dashed and dotted curves) for late-type targets.
To fit the distributions we
used the Gaussian plus constant model for early-type targets, and
the exponential plus constant model for late-type targets, namely,
\begin{equation}
f_E(\Delta v) = f_{E1} {\rm exp}(-\Delta v^2/2\sigma_{\Delta v}^2)+f_{E2},
\end{equation}
and
\begin{equation}
f_L(\Delta v) = f_{L1} {\rm exp}(-\Delta v/\sigma_{\Delta v})+f_{L2},
\end{equation}
respectively. We obtained the best-fit values $\sigma_{\Delta v} = 
269\pm10$ (E-e), $229\pm27$ (E-l), $185\pm 18$ (L-e), and $125\pm 15$ km s$^{-1}$
(L-l). 
Our choices of the velocity limits $\Delta v_{\rm max}=600$ km s$^{-1}$ for early-type
targets and 400 km s$^{-1}$ for late types, take into account this dependence
of pairwise velocity on morphology. 

$\Delta v_{\rm max}$ should also depend on
luminosity, but our choice is a conservative one for galaxies brighter than
$M_r = -19.5$ and it is expected that the dependence is not strong. 
For example, 
if we extend the Faber-Jackson (Faber \& Jackson 1976) or 
Tully-Fisher relations (Tully \& Fisher 1977)  to the kinematics 
of galaxy pairs and adopt a constant mass-to-light ratio, 
we obtain $\Delta v \propto 10^{-0.1 M_r}$.
So there will be 26\% difference in $\sigma_{\Delta v}$ on average for
galaxies with one magnitude difference.
For target galaxies brighter than $M_r=-20.5$ the Gaussian fit gives
$\sigma_{\Delta v}=$ 326 (E-e), 219 (E-l), 224 (L-e), and 152 (L-l)
km s$^{-1}$, for the four cases,
and the ratios of $\sigma_{\Delta v}$ between early and late-type target galaxies
are 326/224 and 219/152, both close to 1.4.

%When the velocity distributions for late-type targets are also 
%fit by Gaussian, we obtain
%$\sigma_{\Delta v}=177$ (L-e) and 123 (L-l), so the velocity dispersion of galaxies 
%around early-type targets  is 1.5 and 1.9 times larger than that around late-type
%targets for early- and late-type neighbors, respectively. 
Besides the morphology and luminosity
dependence the velocity difference is also a function of $r_p$.
For the target galaxies with $-19.5>M_r>-20.5$ the neighbor galaxies
located at $r_p = 100-300 h^{-1}$ kpc have velocity dispersions of
257, 153, 205, and 147 km s$^{-1}$ for the four cases, respectively.
The velocity dispersion becomes lower for early-type targets, but higher for 
late-type targets, and correspondingly, the ratio becomes smaller (1.0--1.2).

The relation between the velocity dispersion and the virial mass, 
$M_{\rm vir} \propto \sigma_{\Delta v}^{\beta}$, is needed to convert 
the measured velocity difference dispersion to the halo mass.
Dark matter halo virial relation estimated from simulated $\Lambda$CDM 
models yields $\beta\approx 3.0$ (Evrard et al. 2008).
An NFW halo with no or slightly anisotropic velocities have 
$\beta\approx 2.5$ when the dispersion is measured at $r_p=100 h^{-1}$kpc 
(Conroy et al. 2007), and $\beta = 2.0$ when the dispersion is measured 
at $r_p = 250 h^{-1}$kpc (Prada et al. 2003).
Figure 2 tells that the ratio of $\sigma_{\Delta v}$ for early and late-type 
targets is about 1.4 for $r_p <100 h^{-1}$kpc.
If we adopt $\beta=2.5$, the ratio of dark halo virial mass for early and
late-type targets will be about 2.
This is why we adopted $\gamma({\rm early}) = 2 \gamma({\rm late})$.
Our main conclusions are essentially independent of this approximation.

%A measurement of the slope in the relation
%between the velocity dispersion and the virial mass, $M_{\rm vir}
%\propto \sigma_{\Delta v}^{\beta}$ using SDSS bright galaxies, 
%gives $\beta\approx 2$ at 350 $h^{-1}$kpc from the central halo 
%where is near the virial radius (Prada et al. 2003) even though
%$\beta \approx 3$ at smaller scales (REFF).
%It implies that the early-type galaxies are typically twice more massive 
%than the late types at the same luminosity when $\sigma_{\Delta v}$
%is about 1.4 times larger for early types. 
%This is why we adopted $\gamma({\rm early}) = 2 \gamma({\rm late})$.

\section{Results}
In this section we study the dependence of 
galaxy properties on three environmental factors; the nearest neighbor distance 
$r_p$, nearest neighbor's morphology and the large-scale background density 
$\rho_{20}$.  We are going to consider 
the physical parameters like morphology, $r$-band luminosity,
$u-r$ color, $g-i$ color gradient, equivalent width of the $H\alpha$ line,
$i$-band concentration index, central velocity dispersion, and $i$-band
Petrosian radius. These parameters reflect most of major physical properties
of galaxies from morphology and mass to internal kinematics and 
SF activity.

In most cases we fix the absolute magnitude of target galaxies in a narrow
range between $-19.5$ and $-20.5$ to examine the pure environmental
effects with the effects due to the coupling of a parameter
with luminosity taken out.
We have also studied a set of brighter target galaxies with $-20.5>M_r>-21.5$, and
obtained qualitatively the same results but with much worse statistics.
All late-type target galaxies with the $i$-band isophotal axis ratio less than
0.6 are discarded in the following analysis whenever necessary in order to
reduce the wrong trends that can be produced by the internal extinction
and by the corresponding dispersion in luminosity
(Choi et al. 2007).

\subsection{Morphology}

\begin{figure} 
\center
%\plotone{fig3.eps}
\includegraphics[scale=0.5]{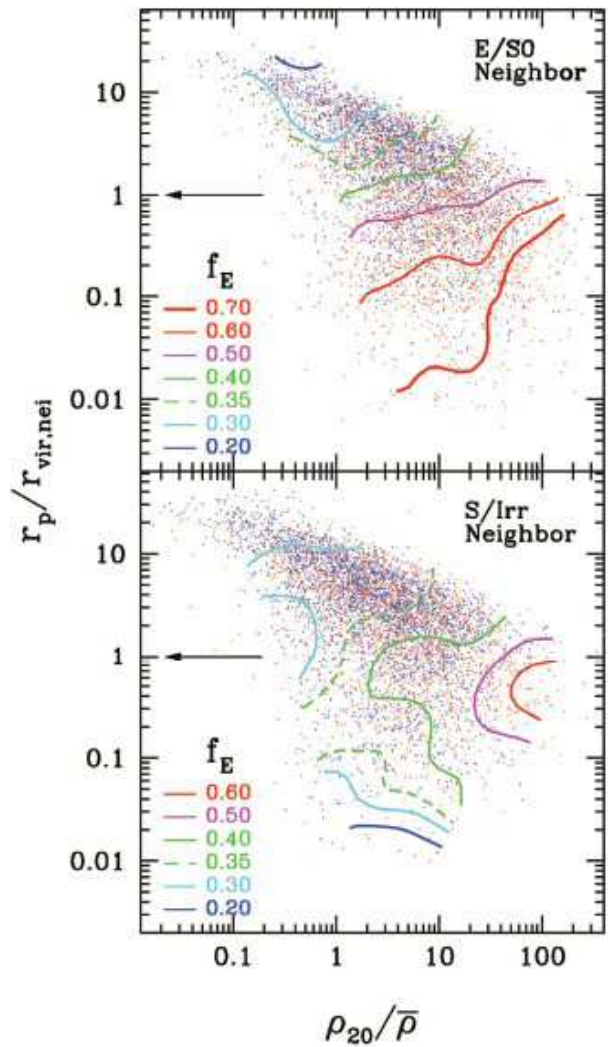}
\caption{(upper) Morphology-environment relation when the nearest neighbor 
galaxy is an early-type galaxy.  Red points are early-type target 
galaxies, and blue points are late-type target galaxies.
Absolute magnitude of galaxies is fixed to a narrow range of $-19.5>M_r>-20.0$. 
Contours show constant early-type galaxy fraction $f_E$.  Contours are 
limited to regions with statistical significance above $1 \sigma$.
(lower) Same, but for the late-type neighbor case.
Arrows at $r_p/r_{\rm vir,nei} = 1$ are drawn to guide the eye.
}
\end{figure}

Park et al. (2008) made an extensive study of the dependence of galaxy
morphology on the density $\rho_n$ attributed 
to the nearest neighbor and also on neighbor's morphology. 
Dependence on the large-scale background density $\rho_{20}$ was also studied 
by analyzing subsets of galaxies located in three different bins of $\rho_{20}$.
It was found that, when luminosity is fixed, the probability for a randomly 
chosen target galaxy to have an early-type, $f_E$, depends mostly on
the projected separation $r_p$ 
when $r_p > r_{\rm vir,nei}$, the virial radius of the nearest neighbor galaxy.
But when $r_p< r_{\rm vir,nei}$, $f_E$ depends on all three environmental factors
in the sense that $f_E$ is an increasing function of both $\rho_n$ and $\rho_{20}$
when the neighbor is an early-type galaxy, but that $f_E$ first increases and then 
decreases as the galaxy approaches a late-type neighbor while it is still
an increasing function of $\rho_{20}$
(see Figure 3 of Park et al. 2008).

To clarify the morphology dependence on those environmental factors we inspect
the behaviour of $f_E$ in the $r_p$-$\rho_{20}$ parameter space in Figure 3. 
The absolute magnitude of target galaxies is limited to a narrower bin,
$-19.5>M_r>-20.0$, in order to reduce the effects of the correlation
of morphology and density with luminosity. 
Figure 20 of Park et al. (2007) showed the same kind of plot, but 
neighbor's morphology was not differentiated there.

The upper panel of Figure 3 shows the dependence of $f_E$ on $r_p$ and $\rho_{20}$
when the nearest neighbor is an early-type galaxy. The lower panel is
the case when the neighbor is a late-type galaxy. 
Since the mean separation between galaxies decreases as the background density
increases, there is a correlation between $r_p$ and $\rho_{20}$.
Because of this statistical correlation galaxies are distributed along
the diagonal in the figure. But there is a large dispersion in $r_p$
at a given $\rho_{20}$, particularly when $\rho_{20}\ga\bar{\rho}$
or $r_p \la r_{\rm vir,nei}$. 
For example, at a fixed large-scale background  density of $\rho_{20}=  
10{\bar\rho}$ the separation to the nearest neighbor of an early-type galaxy 
with $M_r=-20$ can be as small as 0.003 $h^{-1}$Mpc or as large as 2 $h^{-1}$Mpc. 
We will see a large variation in galaxy properties as $r_p$
changes while $\rho_{20}$ is fixed.
When $\rho_{20}$ is very small, however, the correlation is tight and
the information in $\rho_{20}$ and $r_p$ is rather redundant.
When $\rho_{20}$ is higher than $10\bar{\rho}$, most tight pairs with
$r_p \la 0.1 r_{\rm vir,nei}$ are early-type galaxies.

Smooth distributions of $f_{E}$ are obtained from the ratio of the sum
of weighted number of early types to the sum of weighted number of all galaxies
within the smoothing kernel at each point of the parameter space. 
A fixed-size spline-kernel is used to give the weights.
Contours denote the constant levels of $f_E$ and are 
limited to regions with statistical significance above $1 \sigma$.

A striking difference between two contour plots is seen where 
$r_p  < r_{\rm vir,nei}$, namely when the target galaxy is located inside
the virial radius of its neighbor. If the neighbor is an early-type galaxy,
the contours have positive slopes there, meaning that $f_E$ is increasing for
decreasing $r_p$ and increasing $\rho_{20}$. But if the neighbor is a late-type
galaxy, the contours have negative slopes, telling that $f_E$ decreases for
decreasing $r_p$ even though it is still an increasing function of $\rho_{20}$.
At a fixed $\rho_{20}$, $f_E$  has a maximum at $r_p \sim r_{\rm vir,nei}/3$
for the late-type neighbor case. 
This scale corresponds to $70 - 80 h^{-1}$kpc for the late types
in our sample. Galaxies having their late-type neighbors within this 
critical distance, start to have significant hydrodynamic effects
from neighbor.  Sensitivity of $f_E$ to $\rho_{20}$ exists mainly 
within the virialized region
with $r_p < r_{\rm vir,nei}$. 

When $r_p > r_{\rm vir,nei}$ both panels show that galaxy morphology
depends mostly on $r_p$ and is nearly independent of background density
$\rho_{20}$ as can be noticed from nearly horizontal contours.
Contours at $r_p >r_{\rm vir,nei}$ are slightly contaminated by the trend
that the galaxies at the upper edge of the distribution, having the 
largest $r_p$
at a given $\rho_{20}$, are relatively brighter and tend to be earlier in type
(see Fig. 5 below). 
As we decrease the absolute magnitude bin size, the contours become flatter,
and show weaker dependences on $\rho_{20}$ in this  pair separation range.
But a slight dependence of $f_{E}$ on $\rho_{20}$ seems persisting at 
$r_p > r_{\rm vir,nei}$ for the early-type neighbor case.

Figure 3 confirms Park et al.(2008)'s findings that the effects of the nearest
neighbors are critically important to galaxy morphology and 
that the large-scale density matters
only when pairs are closer than the virial radius.

The early-type fraction can be very high in very high large-scale density 
regions even when the neighbor is a late type. On the other hand,
for the isolated galaxies with $r_p \gg r_{\rm vir,nei}$ located in very low
density regions ($\rho_{20}\ll \bar{\rho}$), the early-type fraction \fe\
asymptotically approaches about 0.2, which might be the inborn morphology 
fraction because the galaxy interaction and merger rates are low there.
Gott \& Thuan (1976) proposed that galaxy morphology is
determined by the amount of gas left over at maximum collapse
of the protogalaxy. Primordial elliptical galaxies are expected to form 
if star formation is finished by the time of maximum collapse.
This can happen if the star formation time scale is shorter than the
collapse time, which is more likely in high density regions.
In reality, according to simulations galaxies continously accrete gas and other objects
till the present time, and it is difficult for SF to end in the course of formation.

It is also conceivable that an isolated system of late-type galaxies is formed
in low density regions and transforms to an isolate early-type galaxy 
by consuming all cold gas in the system
through a series of close interactions and mergers (Park et al. 2008).
These isolated early types cannot be formed by the mechanisms like 
strangulation because they have no nearby larger halo. 
This kind of early types is analogous to the central dominant 
elliptical galaxies in the `fossil groups' which is the end-result
of galaxy mergers (Jones et al. 2003; Ulmer et al. 2005; Mendes de Oliveira 
et al. 2006; D'Onghia et al. 2005; Adami, Susseil, \& Durret 2007;
von Benda-Beckmann et al. 2008 and references therein). 
%This may be because of neighbors other than the nearest one, which are
%also likely to be early types.
On the other hand, a merged object can become an early-type galaxy due to the
AGN heating (Croton \& Farrar 2008) and remain isolated.

\subsection{Luminosity}

\begin{figure} 
\center
%\plotone{fig4.eps}
\includegraphics[scale=0.45]{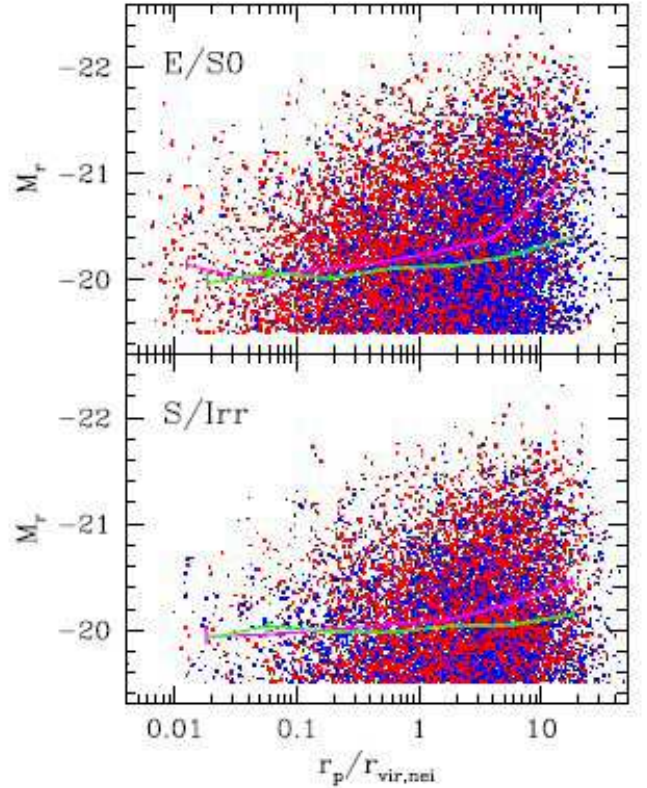}
\caption{Dependence of absolute magnitude of galaxies on morphology and
neighbor separation. The upper panel is for the early-type target galaxies,
and the lower panel is for late types. All galaxies brighter than $M_r = -19.5$
are plotted. The median $M_r$ relations are drawn for early-type neighbor 
(red dots, magenta line) and late-type neighbor (blue dots, green line) cases.
}
\end{figure}

\begin{figure*} 
\center
%\plotone{fig5.eps}
\includegraphics[scale=0.6]{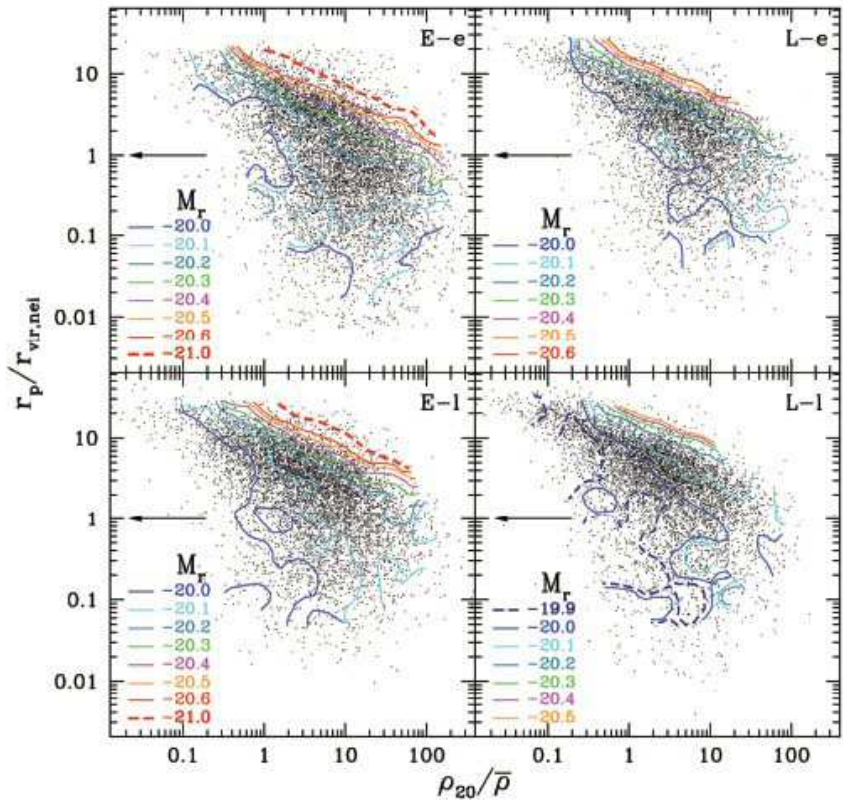}
\caption{The median absolute magnitude contours in the projected pair separation 
$r_{p}$ versus the large-scale background density $\rho_{20}$
space. Points are all target galaxies brighter than $M_r = -19.5$. 
Four cases are given, the early-type target galaxies having an early-type neighbor
(E-e), early-type targets having a late-type neighbor (E-l), late-type targets
having an early-type neighbor (L-e), late-type targets having a late-type 
neighbor (L-l).}
\end{figure*}
Figure 4 shows the $r$-band absolute magnitude of galaxies
as a function of $r_p$. 
The lines are the median relations.
We use all target galaxies with $M_r < -19.5$ in this section.
Late types with the axis ratio $b/a<0.6$ are still discarded because
their absolute magnitudes are not reliable.
As Figure 6 of Park et al. (2008) showed,
more isolated galaxies are brighter on average and such trend is greater for
early-type galaxies. Figure 4 also shows that galaxies with an early-type 
neighbor are much brighter than those with a late-type neighbor when they have
relatively large separations from neighbor ($r_p > r_{\rm vir,nei}$).

Figure 5 shows that the dependence of $M_r$ on $r_p$ and $\rho_{20}$.
Smooth distributions of $M_r$ are found by the following method 
(it is applied to all panels in Figs. 5, 7, and 9).
At each location of the $r{_p}/r_{\rm vir,nei}$-$\rho_{20}/\bar{\rho}$ space,
where a smoothed median value is to be estimated, we first sort 
the parameter values of the galaxies contained within a certain radius
from the location. In the case of the tophat smoothing, these galaxies get a
uniform weight and the median would be the value of the $N/2$-th galaxy
when the total number of galaxies within the smoothing radius is $N$.
We adopt a spline kernel smoothing for more accurate results.
We assign a spline kernel weight 
$w_i$ to each galaxy within the smoothing radius.
The median is given by the parameter value of the $\Sigma w_i /2$-th
galaxy in the sorted list, 
where $\Sigma w_i$ is the sum of all weights given to the galaxies
within the smoothing kernel.
The median represents the typical value of a physical parameter more reliably 
than the mean, when the distribution is very skewed as in the case of 
$W(H\alpha)$ in particular.

Figure 5 shows that there is strong environmental dependence of $M_r$ 
at separations $r_p \ga r_{\rm vir,nei}$ while such
dependence fades away at smaller separations. 
At $r_p \la r_{\rm vir,nei}$ the absolute magnitude fluctuates within 0.1
magnitude in all panels without showing significant dependence on $r_p$
or $\rho_{20}$. One should note that a single horizontal or vertical contour
does not show any environmental dependence. There should be a significant
gradient in $M_r$ revealed by a series of parallel contours with different levels 
in order to claim a dependence. 
At a given large-scale density $\rho_{20}$
the brightest galaxies are those having the largest $r_p/r_{\rm vir,nei}$,
namely the most isolated ones. 
The $r_p$-dependence of luminosity persists from very high
density regions to well-inside the voids.
The void galaxies are also participating in luminosity evolution,
but the speed of the evolution is slow because of fewer neighbors. 
%Correspondingly, the maximum luminosity that galaxies can reach is relatively lower 
%in under-dense regions than that of galaxies in high density regions.
%The upper left part of each panel of Figure 5 shows the trend.

A fact noted  from Figure 5 
is that such a trend exists in all density environments,
all the way from voids to very high density regions with $\rho_{20}/{\bar\rho} 
\sim 100$ (but note that we cannot resolve the cluster regions where
$\rho_{20}$ itself is higher than the virialized density). Figure 5 also shows that the 
luminosity-density relation (the horizontal direction in Fig. 5) is a very
strong function of neighbor distance. 
For example, in the case of the E-e galaxies, 
luminosity of very isolated galaxies
with $r_p\approx 20 r_{\rm vir,nei}$ rises quickly as $\rho_{20}$ increases
and reaches $M_r=-21.0$ at the density 
as low as $\rho_{20}\approx{\bar\rho}$.
But for galaxies with  $r_p\approx 2 r_{\rm vir,nei}$ it rises much slowly
for increasing $\rho_{20}$ and reaches $M_r=-21.0$ only  at 
$\rho_{20}\approx 100 {\bar\rho}$. The typical luminosity of the galaxies 
with neighbors at much closer distances never reaches this magnitude.

It can be also noted from Figure 5 that the early-type galaxies tend not to 
have late-type neighbors at very small distances. There is almost 
no E-l galaxy at $r_p<0.02 r_{\rm vir,nei}$ 
while many E-e galaxies have close neighbors at these separations 
(compare the scatter plots in the left column of Fig. 5).
This may be because the early-type galaxies having their late-type neighbors
at $r_p<0.02 r_{\rm vir,nei}$ can acquire enough cold gas and transform
their own morphology to late type. In fact, there are more such tight L-l
galaxies than E-l galaxies as can be seen in the two bottom panels of Figure 5.
Evidence for the morphology transformation from an early to a late type has been
presented by Park et  al. (2008).

\subsection{$u-r$ color}

\begin{figure*} 
\center
%\plotone{fig6.eps}
\includegraphics[scale=0.65]{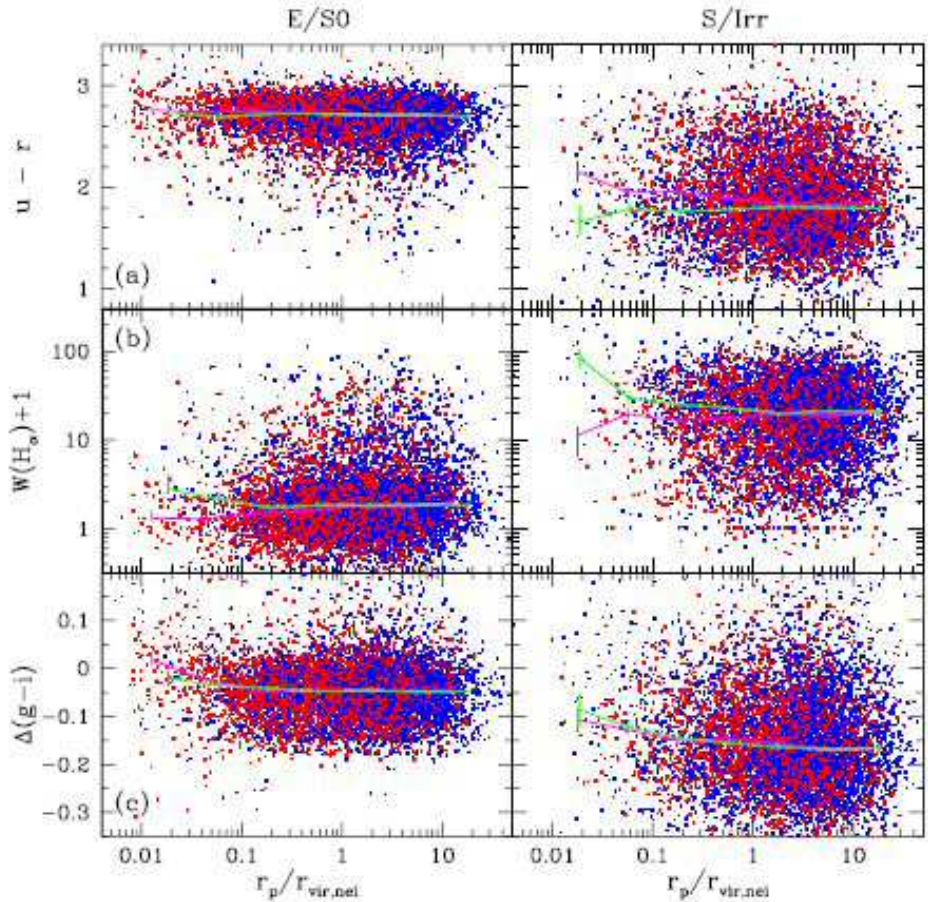}
\caption{Dependence of physical parameters of our target galaxies with 
$-19.5>M_r >-20.5$ on the distance to the nearest neighbor normalized to the
virial radius of the neighbor.
Left panels are for early-type target galaxies and right panels are for late types.
The physical parameters considered here are the $u-r$ color, equivalent
width of $H\alpha$ line, and $g-i$ color gradient.
Median curves are drawn for the cases of early-type neighbor (magenta line, red dots)
and of late-type neighbor (green line, blue dots).
The $r_p$-space is uniformly binned in the logarithmic scale, and in each
bin the median $r_p$ of the galaxies belonging to the bin is used 
for the median curve.
}
\end{figure*}

Figure 6 shows variations of three physical parameters of our target galaxies
with $-19.5>M_r>-20.5$
as a function of pair separation normalized to the virial radius of their
nearest neighbor. Left panels are for the early-type target galaxies,
and right panels for the late types. In each panel
red points are galaxies with early-type neighbors, and blue points are
those with late-type neighbors. Their median values are shown by magenta
and green lines, respectively.

Top left panel shows that the $u-r$ color of early-type galaxies hardly
changes as they approach neighbors. Even when the neighbor is
a late-type galaxy, the change is very small, which is surprising 
because morphology of target galaxies
is strongly affected by the neighbor distance and tends to become a late type
when the neighbor is a late type. 
The main reason for this is, of course, because the subsample is 
already restricted to early types whose color shows a small dispersion. 
But our subsamples are not simply divided by color, 
but divided according to morphology as much as possible (Park \& Choi 2005) 
and correspondingly the early-type subsample contains many blue galaxies. 
For example, among the early types plotted in Figure 6 
more than 7\% have $u-r$ color bluer than 2.4. One could expect to
see some blueing trend for early types interacting with a late type.
Even though the total color does not change much,
we will see below that the SF activity of early-type galaxies 
having late-type neighbors is actually enhanced when the separation
is much smaller than the virial radius, suggesting cold gas transfer from
their neighbors.

% It may be also because late-type morphology
%is assigned to an originally early-type galaxy when its color is significantly
%changed during the interactions.

The upper right panel of Figure 6 shows that,
when the target galaxy is a late type,
its $u-r$ color clearly becomes redder as it approaches an early-type
neighbor, but does not change much if the neighbor is 
a late type. The bifurcation occurs at $r_p\sim r_{\rm vir,nei}$.
From Figure 6 one can understand why it has been so difficult to detect
the change in color for interacting pairs (de Propris et al. 2005). 
Early types do not show any significant
changes, and the changes in late types depend on neighbor's morphology.
Without dividing the sample according to target's and neighbor's morphology
one would find little change in color.

In the left column of 
Figure 7 the behaviour of the median $u-r$ color in the $r_p$-$\rho_{20}$ 
space is inspected.
We again find that the color of early types
is very insensitive to both $r_p$ abd $\rho_{20}$. The median 
color of the early-type galaxies having an early-type neighbor (the E-e galaxies)
or a late-type neighbor (the E-l galaxies) varies only about 0.02 in 
the $r_p$-$\rho_{20}$ space.

The bottom two panels of the left column of Figure 7 show the median $u-r$ 
color contours for late-type target galaxies. 
It can be seen that the $u-r$ color of the L-e galaxies depends on $r_p$ 
weakly when $r_p \ga 0.1 r_{\rm vir,nei}$ but strongly for
$r_p \la 0.05 r_{\rm vir,nei}$. 
The dependence is negligible at $r_p > r_{\rm vir,nei}$.
The L-l galaxies tend to be slightly bluer as $r_p$ decreases.
This is why the L-e and L-l cases start to separate from each
other at $r_p \sim r_{\rm vir,nei}$ and then diverge at 
$r_p \la 0.05 r_{\rm vir,nei}$.

Figure 7 also shows that, when $r_p \ga 0.05 r_{\rm vir,nei}$, late-type galaxies 
become redder as their background density becomes higher.
This weak residual dependence of color of late-type galaxies on $\rho_{20}$ 
after fixing luminosity has been also reported by Park et al. (2007).
The residual dependence is weak; the total variation in $u-r$ of late types
is about 0.2 magnitude as $\rho_{20}/{\bar\rho}$ varies from 
0.1 to 100.
If luminosity or morphology are not fixed, the color variation is much larger
(see Fig. 11 of Park et al. 2007).
The large-scale density dependence of $u-r$ is probably a result of 
the accumulated effects of galaxy-galaxy interactions and mergers whose frequency is
higher in higher density regions.

An interesting fact from these plots is that the
dependence of color on $\rho_{20}$ becomes negligible when 
$\rho_{20}<{\bar\rho}$, which can be noticed from widening of contour 
separation and serpentine contours at small $\rho_{20}$.
We will also see in the next section that the SF activity as measured by
$W(H\alpha)$ does not depend on the background density $\rho_{20}$ either
when $\rho_{20} <{\bar \rho}$ and $r_p \ga r_{\rm vir,nei}$.
%These seem to indicate that the dependence of color and SFA on $\rho_{20}$
%is not caused by a direct influence of background environment, but
%is caused by indirect effects through some properties of neighboring galaxies
%like halo gas pressure.

%The left column of Figure 7 shows that early-type galaxies are nearly independent 
%of the large-scale background density in color once luminosity is fixed.
%But the bottom two panels show a residual effect of $\rho_{20}$
%on late-type galaxies. At $r_p \ga 0.05 r_{\rm vir,nei}$, i.e. when galaxies
%are not merging, late-type galaxies become redder by $0.2\sim 0.3$ 
%magnitude in typical $u-r$ color as $\rho_{20}$ changes from 0.1 to 100 times 
%${\bar\rho}$ (see also Fig. 11a of Park et al. 2007).
On the other hand, when
the background density is very high ($\rho_{20}\ga 50{\bar\rho}$),
the color of late-type galaxies appears to depend only on $\rho_{20}$.
The density range corresponds to the cluster environment, and a cluster
acts like a giant early-type galaxy transforming member galaxies
into redder ones.
Recently, Park \& Hwang (2008) have studied properties of galaxies
near and within Abell clusters, and found a characteristic scale of $2 - 3$
times the cluster virial radius across which various galaxy properties 
suddenly start to show dependence on the clustercentric radius.
%Dependence of galaxy color on the clustercentric radius also appears suddenly
%at the characteristic scale. 
Since the cluster galaxies are smeared along the
line of sight due to the finger-of-god effect in our analysis, reddening of late types 
due to clusters is expected to appear rather smoothly as a function of $\rho_{20}$.

%One should note that the overall value and $\rho_{20}$-dependence of color are 
%quite different in two panels with different neighbor morphology. It
%demonstrates importance of neighbor morphology in determining galaxy color.
%The $\rho_{20}$ dependence of late-type galaxies' color is probably caused by
%the hot halos of the galaxies and their neighbors whose properties
%depend on $\rho_{20}$, rather than by an innate relation between
%galaxy properties and environment.

Kauffmann et al. (2004) claimed that the SF history-density correlation 
is sensitive to small-scale density, but that there is no evidence for the SF 
history to depend on large-scale ($>1$ Mpc) density.
We find in Figure 7 that the $u-r$ color as a measure of SF history
is nearly independent of small and large-scale environments for early-type
galaxies, but depends on both for late-type galaxies even at fixed luminosity.

\begin{figure*} 
\center
\plotone{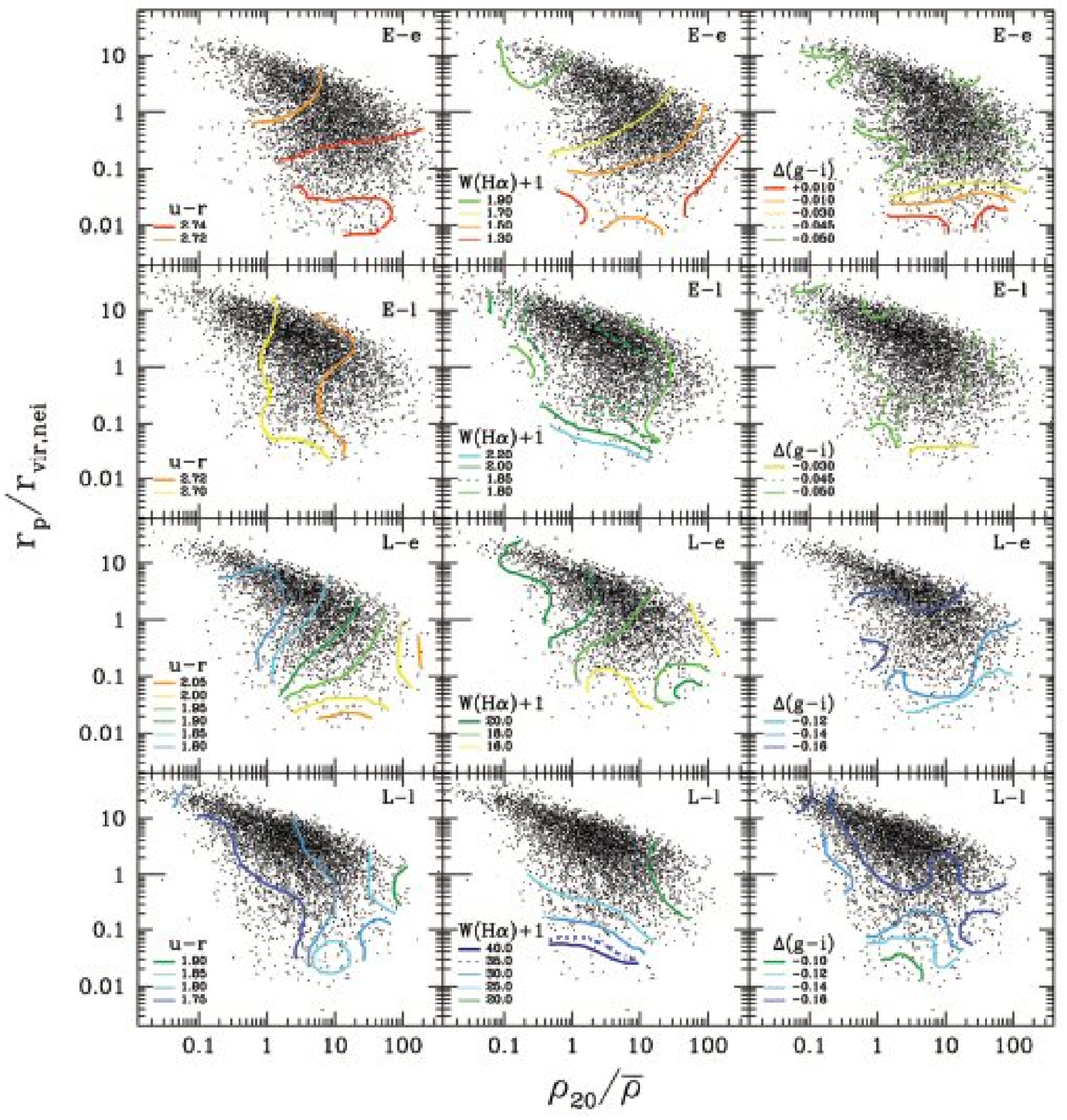}
\caption{Dependence of the $u-r$ color, equivalent width of the $H{\alpha}$ line,
$g-i$ color gradient of galaxies with $-19.5>M_r>-20.5$ on the pair separation 
$r{_p}$ and the large-scale background density $\rho_{20}$.
In each column, galaxies are divided into
four cases, the E-e, E-l, L-e, and L-l galaxies.
Dots are galaxies belonging to each subset.
At each location of the $r_p$-$\rho_{20}$ space the median value of the
physical parameter is found from those of galaxies within a certain 
distance from the location (see section 3.2 for more details).
Curves are the constant-parameter contours.
A short line at $r_p/r_{\rm vir,nei} = 1$ is drawn to guide the eye.
}
\end{figure*}

\subsection{Equivalent width of the $H{\alpha}$ line}

Figure 6b (middle panels) shows variations of the equivalent width of the 
$H\alpha$ line, a measure of the SF activity, as a function of $r_p$ 
for the early-type (left) and late-type (right) target galaxies with $-19.5>M_r>-20.5$.
Galaxies are again distinguished
between those having an early-type neighbor (red dots, magenta line) and
those having a late-type neighbor (blue dots, green line).
Even though the total color of early-type galaxies is not much affected by
interactions, the SF activity measured by $W(H\alpha)$ clearly shows 
dependence on $r_p$ and neighbor's morphology.
This tells us that the color is simply not a very
sensitive SF indicator, especially for massive galaxies.

Early-type galaxies show slightly reduced SF activity when they approach an early-type
neighbor. But the E-l galaxies show relatively stronger 
$H\alpha$ line emission at $r_p < r_{\rm vir,nei}$,
and the gap between two cases becomes wider at $r_p < 0.1 r_{\rm vir,nei}$.

$W(H\alpha)$ of late-type galaxies shows more dramatic variations
as $r_p$ changes. In the right panel $W(H\alpha)$ of the L-e galaxies
is nearly constant down to $r_p\approx 0.05 r_{\rm vir,nei}$ and
then starts to drop below the separation. But $W(H\alpha)$ of the 
L-l galaxies starts to rise at $r_p\approx  r_{\rm vir,nei}$ and
increases rapidly below $r_p \approx 0.05  r_{\rm vir,nei}$.
The two cases start to bifurcate at $r_p\approx  r_{\rm vir,nei}$ and
to diverge at $r_p\approx 0.05  r_{\rm vir,nei}$ as in the $u-r$ case.

We note there are two characteristic scales in the effects of galaxy-galaxy
interactions on the SF activity. 
The first scale is
the virial radius of the nearest neighbor galaxy where difference 
in morphology of the nearest neighbor starts to make the SF activity bifurcate.
The second one is the merger scale, about 0.05 times the virial radius, where
the effects diverge depending on the neighbor morphology.
Since the virial radius of an early or late-type galaxy 
with $M_r = -20$ is 300 or 240  $h^{-1}$ kpc,
at the separation $r_p = 0.05 r_{\rm vir}\approx 12 - 15h^{-1}$ kpc,
a pair of such galaxies should start to physically contact to each other.
%The behaviour of $W(H{\alpha})$ in the $r_p$-$\rho_{20}$ space (middle
%column of Figure 7) indicates that both $r_p$ and $\rho_{20}$ are determining
%factors of SF activity, with the major effects coming from $r_p$.

The middle column of Figure 7 shows dependence of $W(H\alpha)$ on $r_p$ and 
$\rho_{20}$ for four cases.
%The contours represent the smooth field given by the median (not the mean) 
%value of $W(H\alpha)+1$. 
The two top panels indicate that the SF activity of early types 
depends on neighbor's morphology.
The E-l galaxies shows overall higher $W(H\alpha)$ compared to the 
E-e galaxies, and the direction of dependence of $W(H\alpha)$ on
$r_p$ is opposite for the two cases. We also find that $W(H\alpha)$ of 
early-type galaxies decreases slightly as $\rho_{20}$ increases.
The $\rho_{20}$ dependence of $W(H\alpha)$, however, does not always exist,
but exists only over a certain $r_p$ range that depends on the neighbor
morphology.

For late-type galaxies (two bottom panels)
the effects of early- and late-type neighbors are clearly distinguished.
The SF activity of late types is enhanced
significantly as they approach late-type neighbors, but quenched slightly
as they approach early-type neighbors.
In the L-l case the dependence on $\rho_{20}$ at small
$r_p$ is qualitatively similar to the E-l case.
The scale of occurrence of the $r_p$ dependence depends on $\rho_{20}$.
Enhancement of SF activity by late-type neighbors occurs at smaller neighbor
separation in higher $\rho_{20}$ regions.
This can happen if late-type neighbor galaxies tend to be less gas-rich, and 
the cold gas flow from late-type neighbors is less efficient in higher density  regions. 
In all four cases the SF activity does not depend on both $r_p$ and $\rho_{20}$
when galaxies are isolated ($r_p \ga r_{\rm vir,nei}$) and when the
background density is small $\rho_{20} \la {\bar\rho}$.

Balogh et al. (2004b) analyzed the galaxies and groups in the 2dF Galaxy Redshift 
Survey data and the SDSS data, and reported that the fraction of star forming 
galaxies varied strongly with the background density (see also Balogh et al. 2004a).
The signal they found must be mostly due to the correlation of the 
background density with luminosity and morphology. 
They looked at faint late-type galaxies in low
density regions, but bright early types at high densities.
Once luminosity and morphology are fixed, 
the SF activity in galaxies depends very weakly on the background density as
shown in Figure 13c of Park et al. (2007) and in Figure 7 of this work.
Balogh et al. (2004b) also presented evidence that the fraction of $H{\alpha}$ 
emitting galaxies 
is mostly dependent on the small-scale environment at high densities, 
but on the large-scale environment at low densities.  
We found different results. The second column 
of Figure 7 shows that the SF activity of late-type galaxies having a late-type
neighbor depends differently on $r_p$ at different $\rho_{20}$ 
in such a way that in higher density regions the SF activity rises at smaller $r_p$.
But there is only a weak trend that is opposite to this for the L-e galaxies.

Nikolic et al. (2004) claimed that the mean SF activity is 
significantly enhanced for $r_p < 30$ kpc. For late-type targets
the enhancement is found out to 300 kpc
regardless of neighbor's morphology.
We find their results are true only when the neighbor (or target) galaxy 
is a late type.  For the `L-l' galaxies
we find the enhancement of SF activity extends out 
to $r_p \la r_{\rm vir,nei}\sim 300$ kpc.
But when the neighbor is an early type, 
the SF activity drops at $r_p \la 30$kpc or 
at $r_p \la 0.1 r_{\rm vir,nei}$. 
This shows us again that the effects of interaction become manifest
when the sample is split according to the morphology of target and
neighbor galaxies.
They also reported that there is no dependence of the SF activity on 
neighbor morphology nor mass.  
This is certainly not supported by our results.

Alonso et al. (2006) reported that there is a threshold for the SF
activity induced by interactions at $r_p=100h^{-1}$kpc. We do not find evidence
for a threshold at that scale. Our results suggest thresholds only at 
the virial radius and merger scale, which are roughly $\sim300$ and $15 h^{-1}$kpc
for galaxies in our sample, respectively.
They also found that interactions are more effective at triggering SF
activity in low and moderate density environments. This is consistent with
our results only when the neighbor galaxy is a late type. In the second and bottom
panels (the E-l and L-l cases) of Figure 7 one can see $W(H{\alpha})$ starts
to increase at $r_p\la 0.2  r_{\rm vir,nei}$ in very low density regions but 
at  $r_p\la 0.05 r_{\rm vir,nei}$ in high density regions.
The influence of early-type neighbors is weaker in low density regions,
and the SF activity is less suppressed.

%At a given pair separation the SFR is always 
%lower at higher $\rho_{20}$.
%This trend is weak, but exists not only in high density regions, but also
%in intermediate and low density regions (see Figure 13c of Park et al. 2007
%for a comparison), and 

\subsection{$g-i$ color gradient}

We use the $g-i$ color difference between the central region with $R<0.5R_{\rm Pet}$
and the annulus with $0.5R_{\rm Pet} < R < R_{\rm Pet}$ as a measure of color
gradient. Here $R_{\rm Pet}$ is the Petrosian radius (Petrosian 1976; 
Blanton et al. 2001) in the $i$-band. 
The difference is made in
such a way that positive $\Delta(g-i)$ means a bluer central region relative 
to the outer region.
We corrected $\Delta(g-i)$ for the inclination and seeing effects as described 
by Park \& Choi (2005) and Choi et al. (2007).
We use the gradient in $g-i$ color rather than $u-i$ color because the $u$-band 
surface photometry is noisy for some galaxies.
But the surface photometry of SDSS galaxies in the $g$, and $i$-bands
can be done relatively accurately for those in the spectroscopic sample 
(apparent $r$-band magnitude $m_r < 17.6$).

The bottom panels of Figure 6 clearly show that the central region of galaxies
undergoing interactions and mergers becomes bluer relative to the outside.
For early-type target galaxies
the effects are manifest only well within the virial radius. 
But for late types the effects start to appear at separations 
larger than $r_{\rm vir,nei}$.
Difference due to different neighbor morphology is small, which might seem 
inconsistent with the trends seen for $u-r$ and $W(H{\alpha})$.

If the dependence of color and SF activity on neighbor's morphology
is because of difference in the influence of cold and hot gases of the neighbor galaxy,
one might think the color gradient depends on $r_p$  also differently for different
neighbor morphology.
But Figure 6 shows the color gradient always increases for decreasing $r_p$ 
independently of the morphology of target and neighbor galaxies.
It means that, as early-type galaxies approach their neighbors, 
the central part becomes bluer but the outer part becomes redder,
making the color gradient increase but their total color remain almost the same.
For late-type galaxies the center becomes bluer for both early
and late-type neighbor cases, but the outer part becomes much redder
for the early-type neighbor case, making the total color redder.

Early-type galaxies must be significantly reducing the SF activity 
in the outer part of their neighboring late-type galaxies.
It is known that the SF activity of late-type galaxies in clusters is severely
reduced in the outer disk, with normal or enhanced activity
in the inner disk (Boselli \& Gavazzi 2006). In other words, the color gradient
of late types becomes more positive (redder outside) in the cluster environment. 
Quenched SF activity in cluster spirals
is often explained by the gas depletion through hydrodynamic interactions
with the hot intracluster medium such as the ram pressure stripping, viscous
stripping, thermal evaporation, and strangulation (cf. Boselli \& Gavazzi 2006
and references therein).
We now find in Figure 6 that late-type galaxies in general environment 
experience very similar changes in the SF activity. In particular,
when they approach an early-type galaxy (instead of a cluster), they show
redder total color, reduced SF activity, and more positive color gradient.

%Figure 7 (the panels in the third row) shows that,
%in the very high density environment with
%$\rho_{20} \ga 30 {\bar\rho}$, the L-e galaxies have redder color and more
%positive color gradient compared to those in lower density regions.
%One might think it is the hot intracluster medium that is responsible for this.
%However, there is evidence against this idea.
%The bottom right panel of Figure 7 shows that the L-l galaxies located
%in high density regions do not show increased color gradient.
%The color gradient of late-type galaxies may be modified if
%there is a significant mass transfer from its neighbor.
%But Figure 7 says the color gradient does not increase in high density regions
%even when the late types are not very close to their late-type neighbor
%(i.e. no increase of $\Delta(g-i)$ at $r_p > 0.1 r_{\rm vir,nei}$
%in high density regions).
%This means that the intergalactic medium in the relatively high density regions with
%$30 \la \rho_{20}/{\bar\rho} \la 200$ cannot directly quench the SF activity in
%late-type galaxies. These regions correspond to galaxy groups or 
%the outer part (near the cluster virial radius) of galaxy clusters.

It is important to note that the SF quenching phenomenon is now found for late-type 
galaxies outside the cluster environment (see the discussion section below
for possible mechanisms).
Figure 7 shows that an approach to an early-type neighbor
makes a late-type galaxy redder in color, weaker in $W(H\alpha)$ and 
more positive in color gradient in any background density environment
(see the panels in the third row of Fig. 7).

\subsection{Concentration}

\begin{figure*}
\center
%\plotone{fig8.eps}
\includegraphics[scale=0.65]{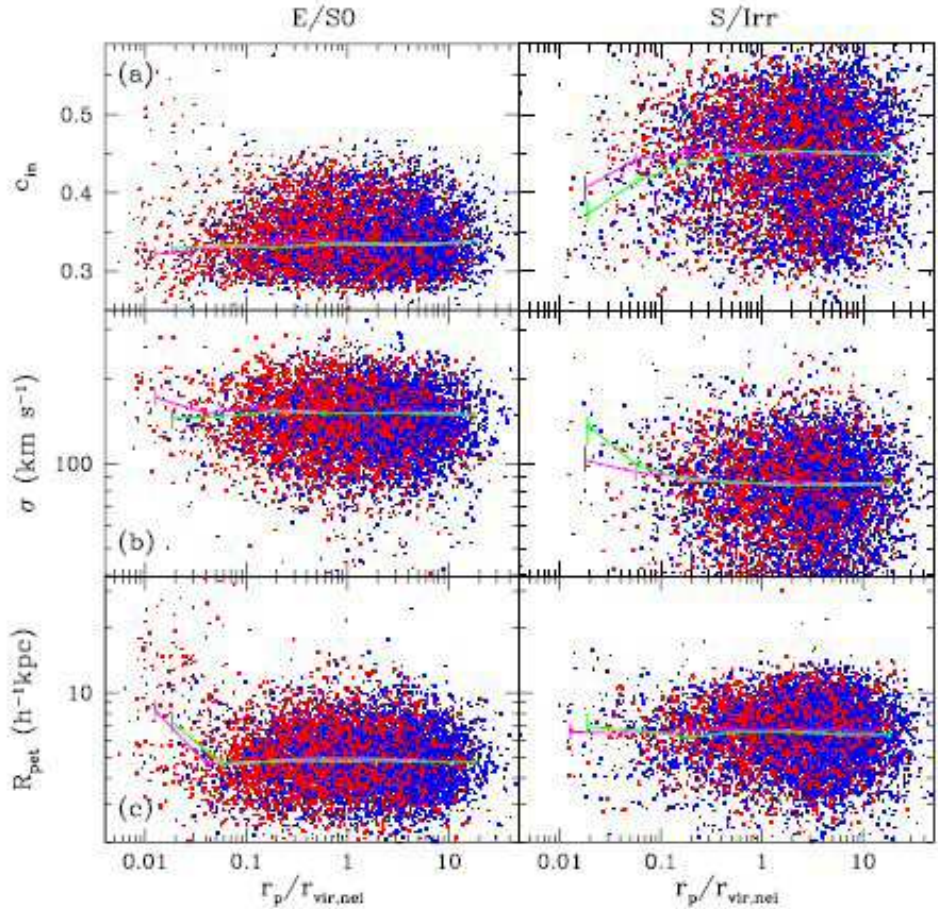}
\caption{Dependence of the physical parameters of our target galaxies
with $-19.5>M_r >-20.5$ on the separation between the target galaxies and
their nearest neighbor galaxy. The physical parameters considered here are
the inverse concentration index $c_{\rm in}$, central velocity dispersion $\sigma$,
and Petrosian radius $R_{\rm Pet}$. Left panels are for early types, and right panels 
for late types. Cases are further divided into early-type neighbor (magenta median
curves, red dots) and late-type neighbor (green curves, blue dots) cases.
}
\end{figure*}

\begin{figure*}
\center
\plotone{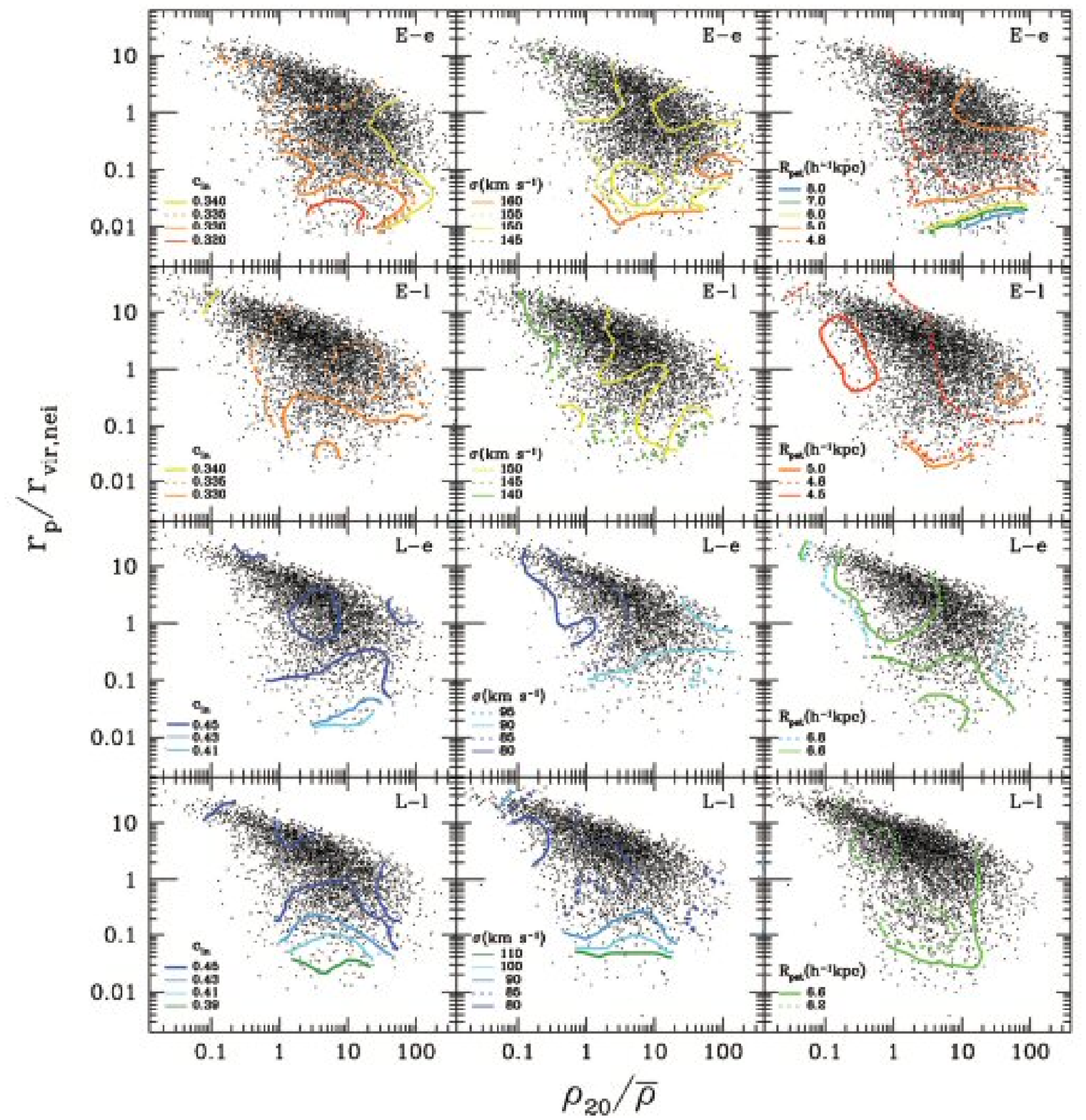}
\caption{The median contours of the inverse concentration index $c_{\rm in}$,
central velocity dispersion $\sigma$, and Petrosian radius $R_{\rm pet}$ in
the neighbor separation $r_p / r_{\rm vir,nei}$ versus the large-scale density
$\rho_{20}/\bar{\rho}$ plane.
Cases are divided into four cases; the E-e (top panels), E-l (second panels),
L-e (third panels), and L-l (bottom panels) galaxies. Galaxy size as measured 
by the Petrosian radius has a potential systematic effect at small $r_p$. 
Most contours look noisy except when $r_p \la 0.1 r_{\rm vir,nei}$ in some
panels. This is mainly because galaxy properties are
almost independent of both $r_p$ and $\rho_{20}$ when galaxy luminosity
and morphology are fixed, and not because statistical uncertainties are large.
}
\end{figure*}

We adopt the inverse concentration index $c_{\rm in}$ to quantify the radial
surface brightness profile of galaxies. It is defined by $R_{50}/R_{90}$ where 
$R_{50}$ and $R_{90}$ are the semi-major axis lengths of ellipses
containing 50\% and 90\% of the Petrosian flux in the $i$-band, respectively,
and is corrected for the seeing effects (Park \& Choi 2005).

Top panels of Figure 8 show the dependence of $c_{\rm in}$ on the pair separation.
Early-type galaxies show smaller change in concentration as they 
approach neighbors probably because they are tidally more stable due to smaller size
and compactness, and also because the tidal energy deposit is relatively
smaller as the relative velocity with the neighbor is higher for early types
than for late types (see Fig. 2). Figure 8 tells that galaxies become
more concentrated as they approach their neighbors.
But we note a very slight tendency that
$c_{\rm in}$ of the E-e galaxies first increases 
and then decreases as $r_p$ decreases.  When the galaxy undergoes a merger at
$r_p \approx 0.05 r_{\rm vir,nei}$, the dispersion in $c_{\rm in}$ becomes 
large. When galaxies are merging,
some fraction of mass escapes from them and form tidal
tails and bridges. This makes the Petrosian radius usually larger
and uncertain. Likewise, $c_{in}$ becomes uncertain too.

The tendency that galaxies become more concentrated as they 
approach their neighbors inside the virial radius is much stronger for late types. 
This may be because late types  are tidally more
vulnerable due to their larger size and lower concentration than early types
and because the velocity difference of late-type target galaxies with their
neighbors is relatively smaller and correspondingly the tidal energy deposit
is larger.
The effects are even stronger for the late-type neighbor case because the velocity 
difference between the target and neighbor is even lower in this case
as shown in Figure 2.  More discussion is given in section 4.

A closer look at the variation of $c_{\rm in}$ reveals that $c_{\rm in}$ 
of late types actually first increases and then decreases 
as $r_p$ decreases, but the increase is
very small and the scale of the maximum $c_{\rm in}$ 
(least concentration) differs for different neighbor
morphology ($r_p \approx 0.5$ and $2 r_{\rm vir,nei}$ or $\sim150$ and
$\sim500 h^{-1}$Mpc for early and late-type neighbor cases, respectively). 
This compares with Nikolic et al. (2004)'s result that
$c_{\rm in}$ peaks at $r_p \approx 50 h^{-1}$kpc when morphological types of 
interacting galaxies are ignored. The typical value of $c_{\rm in}$ 
is quite different for early and late-type galaxies, and the mean morphology
of interacting galaxies varies depending on the pair separation. Therefore, if
the morphological type of the target galaxies is not distinguished,
in addition to the genuine trend caused by interaction,
one will also see an apparent trend in $c_{\rm in}$ caused by the change in
the average morphology of the target galaxy as we change $r_p$.

The distribution of $c_{\rm in}$ in the $r_p$-$\rho_{20}$ space 
shown in the left column of Figure 9,
confirms the very weak decline of $c_{\rm in}$ for decreasing $r_p$ in the case of
early-type galaxies (with a weak maximum for the E-e case), 
and a strong decrease for decreasing $r_p$ in the case of late-type galaxies. 
Most contours in Figure 7 and 9 below look noisy particularly when 
$r_p \ga 0.1 r_{\rm vir,nei}$. This is mainly because galaxy properties are
almost independent of both $r_p$ and $\rho_{20}$ when galaxy luminosity
and morphology are fixed, and not because statistical uncertainties are large.

%The general tidal-force background, which should be a monotonic function of 
%$\rho_{20}$, does not seem to directly affect the concentration of galaxies. 
When $r_p > 0.5 r_{\rm vir,nei}$, $c_{\rm in}$ varies within 0.005
in all cases. When $r_p < 0.2 r_{\rm vir,nei}$, $c_{\rm in}$ increases
weakly with $\rho_{20}$ for the E-e galaxies, but does not show a dependence
on $\rho_{20}$ for the E-l galaxies.
The fact that $\rho_{20}$ dependence of early-type galaxies depends on neighbor 
morphology and  the fact that the $\rho_{20}$ dependence of the E-e galaxies depends
on the neighbor distance $r_p$, together imply that $\rho_{20}$ is not directly 
affecting $c_{\rm in}$. 
The bottom two panels of Figure 9 shows a strong increases of concentration
for late-type galaxies as $r_p$ decreases. But the dependence of $c_{\rm in}$ 
on $\rho_{20}$ is not clear.
According to Park \& Hwang (2008), $c_{\rm in}$ of late types decreases
as the clustercentric radius becomes less than the cluster virial radius.

Figure 11c of Park et al.
(2007) showed that $c_{\rm in}$ is nearly constant of $\rho_{20}$ for bright 
early-type galaxies but increases very slightly with $\rho_{20}$
for the galaxies much fainter than $M_*$. Note that our target galaxies are basically
$M_\ast$ galaxies for which the background density dependence of
$c_{\rm in}$ is expected to be small.

Our result is consistent with that of Blanton et al. (2005a) who claimed that
structural properties of galaxies are less closely related to galaxy `environment'
than are their masses and SF histories. However, this is true only
for the large-scale background density environment. We found a significant
dependence of galaxy structural parameters ($c_{\rm in}$ in this section 
and the central velocity dispersion $\sigma$ in the next section) on
the environmental factors like neighbor distance and morphology.
van der Wel (2008) has studied the dependence of galaxy morphology and structure
on environment and stellar mass, and concluded that galaxy structure mainly
depends on galaxy mass but morphology mainly depends on environment.
Even when both galaxy luminosity and morphology are fixed, we still find galaxy 
structure depends sensitively on the neighbor environment.

\subsection{Central velocity dispersion}

We use the velocity dispersion measured by an automated spectroscopic pipeline
called {\tt IDLSPEC2D} version 5 (D. J. Schlegel et al. 2008, in
preparation). Galaxy spectra of SDSS galaxies are obtained by optical
fibers with angular radius of $1.5''$.
The central velocity dispersion measurement has been corrected for the
smoothing effects due to the finite size of the optical fiber
(see section 3.1 of Choi et al. 2007).
Taking into account the finite resolution of the spectrographs, we discarded
galaxies with $\sigma< 40$ km s$^{-1}$. 

The middle panels of Figure 8 show variations of $\sigma$ as a function of
$r_p$ for the four cases. Early types hardly change their central velocity
dispersion, which may be again because early types are fast, compact, tightly
bound, and correspondingly are tidally more stable.
The central velocity dispersion of late-type galaxies monotonically increases
as they approach their neighbors within $r_{\rm vir,nei}$.
The increase is stronger when the neighbor is a late type. It is very likely
that this is because the late-type neighbor on average has a smaller
relative velocity (see Fig. 2), and therefore produce more tidal energy
deposit than an early-type neighbor.
Coziol \& Plauchu-Frayn (2007) have recently inspected asymmetries in galaxies pairs,
and concluded that the features are consistent with tidal 
effects produced by companions.

The middle column of Figure 9 shows the dependence of $\sigma$ on both $r_p$
and $\rho_{20}$ for the four cases. In all cases we notice that
$\sigma$ slightly increases as $\rho_{20}$ increases which was also shown
in Figure 13b of Park et al. (2007) for galaxies with $M_r \approx -19.8$
or $-20.4$. Interestingly, the increase mainly occurs in low and
intermediate density regions. 
The top two panels show that $\sigma$ increases at the smallest
$r_p$ at fixed $\rho_{20}$ for the E-e galaxies, but is nearly a constant as $r_p$
decreases for the E-l galaxies. 
In the case of the L-e galaxies, the $r_p$-dependence of $\sigma$ starts to appear
within $r_p \sim r_{\rm vir,nei}$. 
The neighbor dependence is strong for the L-l galaxies, in particular.

\subsection{Size}

We use the Petrosian radius (Graham et al. 2005) as a measure of galaxy size. 
It is measured  from the $i$-band images taking into account inclination and seeing 
(Choi et al. 2007). In the case of late-type target galaxies, we are using only those
with the $i$-band isophotal $b/a$ ratio greater than 0.6. 
The bottom panels of Figure 8 show variations of 
$R_{\rm Pet}$  as a function of the pair separation. A prominent feature in this plot
is that the early-type galaxies appear much larger at 
$r_p \le  0.05 r_{\rm vir,nei}$ while such increase in size is not noticeable
for late-type galaxies. 

The size of early-type galaxies in very close pairs may have been systematically 
overestimated because of blending. When undergoing mergers, early types are 
expected to survive longer than late types because they are relatively more compact
and faster. Furthermore, the neighbor undergoing a merger with an early-type galaxy 
is very likely to be an early type too (Park et al. 2008). It would be difficult to define
a boundary for a tight pair of early-type galaxies with smooth distribution of stars.
Some of early types can be still identified as separate objects even at
 $r_p < 0.01 r_{\rm vir,nei}\sim 3 h^{-1}$kpc when their outer extended envelopes are already 
merged with those of their neighbors, and the size of such galaxies can be
easily overestimated. 
On the other hand, since the pairwise velocity 
is smaller for late types, it is expected that their cores merge relatively quickly 
and that there are relatively fewer late-type pairs with very small separations. 
Size of late-type galaxies is determined by the light from disk which has the boundary
relatively abrupt compared to those of early-type galaxies. 
This may be why there is no very tight late-type pair and why an increase
in $R_{\rm Pet}$ at small $r_p$ is not observed for late types.

The top two panels of the right column of Figure 9 show $R_{\rm Pet}$ in the
$r_p$-$\rho_{20}$ plane for the early-type galaxies. They show that $R_{\rm Pet}$
first slightly decrease between $0.1 \la r_p/r_{\rm vir,nei}<1$ and then increase
at shorter separations. 
The size of the E-e galaxies increases rapidly at $r_p < 0.05 r_{\rm vir,nei}$,
and the actual scale the galaxy size starts to rise depends on $\rho_{20}$. 
The E-e galaxies 
in high density regions appear larger than those in low density regions
when they merge. Except for this merger scales the size of early types
hardly depends on $r_p$ or $\rho_{20}$.
The size of late-type target galaxies, shown in the bottom two panels, 
is nearly independent of both $r_p$ and $\rho_{20}$.
%But it slightly increases for increasing $\rho_{20}$ at  $r_p > r_{\rm vir,nei}$.

Galaxy size depends strongly on luminosity and morphology
as shown by Figure 4 of Choi  et al. (2007). For example, the typical value
of $R_{\rm Pet}$ of early types varies from 3 to $10 h^{-1}$kpc, and that of 
late types varies from 4.5 to $13 h^{-1}$kpc as $M_r$ changes from $-18.5$ to $-21.5$.
Correspondingly, $R_{\rm Pet}$ of a random galaxy varies significantly as $r_p$ 
or $\rho_{20}$ varies because the average morphology and luminosity change too.
But once we fix morphology and luminosity within one magnitude,
$R_{\rm Pet}$ is effectively fixed showing variation less than 0.5 $h^{-1}$kpc 
except for early types undergoing mergers. 
%Galaxy size seems a relatively
%stable quantity against galaxy interactions and environment. 

\section{Discussion}

In the previous section various physical parameters of galaxies
are studied as a function of three environmental factors; 
the nearest neighbor distance, the nearest neighbor's morphology,
and the large-scale background density.
An important finding was that 
the virial radius of the galaxy plus dark halo systems acts as a landmark
where most of the galaxy properties start to be sensitive to both 
nearest neighbor's distance and morphology.
If the SF activity of galaxies in pairs, for example, is enhanced due 
to the internal mass perturbed by the tidal force of the neighbor, it
will be always enhanced by the galaxy-galaxy interactions. 
Contrary to this expectation,  the middle panels of Figure 6 and the second 
column of Figure 7 show that it can be enhanced or suppressed
depending on neighbor galaxy's morphology unless the background density 
is very high.  The fact suggests that
it is neighbor galaxy's cold or hot gas that 
enhances or suppresses the SF activity, respectively,
%The observation informs us that 
%galaxy morphology and SF activity evolve through hydrodynamic interactions
when their separation is less than the virial radius.
On the other hand, the structural parameters changes with $r_p$ in the same 
direction independently of neighbor's morphology.
This tells that galaxies also evolve through gravitational effects
as they approach each other.

Which physical mechanism is responsible for the correlation in
morphology and SF activity between neighboring galaxies?
Why do they suddenly care about neighbor's morphology when
they are closer than the virial radius?
One possible explanation is the primordial origin.
Galaxies close from each other form to have similar morphology and
SF activity since they are located in nearly the same large-scale
environment.
But this scenario can not explain why galaxy properties suddenly start
to change according to the neighbor morphology at the virial radius.
The crossing time of galaxies across the virial radius is of order of
$\sim10^9$yrs, much shorter than the age of the universe.
Even if there existed a primordial correlation among galaxies
in physical properties over the scale of the dark halo virial
radius, such a sharp transition in correlation will be wiped out 
due to the infall of new neighbors in the course of time.

A physical process that can explain such an onset of conformity in morphology
and SF activity at a characteristic separation, is the direct 
hydrodynamic interactions between approaching galaxies.
A galaxy plus dark halo system contains hot halo gas and/or cold disk
gas which are confined within the virial radius of the system.
When a late-type galaxy enters within the virial radius of its early-type 
neighbor, it will start to experience the hot gas pressure in the 
neighbor system's halo.
The physical processes acting in this situation can be ram pressure effects
due to the collision with the hot gas ball of the neighbor.

The thermal evaporation and viscous stripping of the late type's disk gas during
the passage through the hot halo gas, are less able to account for the 
sharp transition because
there is a time-delay for these processes to change galaxy properties
significantly.
Likewise the quenching of SF after a shutoff of cold gas supply
from the halo gas (strangulation) is less likely too because, 
even if the galaxy really loses its halo gas,
there will be a significant time delay for the disk gas to be consumed. 
%after the galaxy crosses the neighbor's virial radius.
If two galaxies are gravitationally bound, however, they will orbit each
other within the virial radius.
Then intense hydrodynamic interactions can occur repeatly many times or continuously 
before they merge, and all above processes can contribute to change 
the properties of the orbiting galaxies.
Since they will remain within the virial radius as they orbit,
the onset of correlation in galaxy properties at the virial radius can be
observed.

According to numerical simulations, interacting galaxies can start to transfer
their mass after they pass the closest approach point even though the actual
results depend critically on the interaction parameters
(Toomre \& Toomre 1972; Mihos \& Hernquist 1994; Wallin \& Stuart 1992).
Then, it is conceivable that an early-type galaxy enters a late-type galaxy's
virial radius, acquires some cold gas with angular momentum enough to form
a disk, and transforms itself to a late type (Park et al. 2008).
This mass transfer may be the reason why a galaxy becomes more likely to
be a late type as it approaches a late-type neighbor galaxy within
the virial radius 
even though it tends to be an early type as it approaches the late-type 
neighbor outside the virial radius
(see Fig. 3 in section 3 and Fig. 3 of Park et al. 2008). 

In addition to the dependence on the nearest neighbor separation,
galaxy properties also show dependence on the large-scale background density.
The dependence is strong only for morphology and luminosity.
%We have checked by limiting the absolute magnitude of neighbor galaxies
%into a narrow bin that the background density dependence of morphology
%does not rise because of the systematic effect that
%neighbors are brighter in higher density regions.
Very interestingly, the background density dependence of morphology
appears clearly only when a galaxy is located inside the virial radius
of its neighbor. The reverse is true for luminosity;
the background density dependence of luminosity can be seen only when a
galaxy is outside the neighbor's virial radius.
If the background density gives direct impacts on galaxy morphology, 
both isolated galaxies and galaxies in pairs should also respect 
the background density. But they do not. 
%Therefore, the impacts are indirect through the nearest neighbor.

Let us consider the reason for the relation between the large-scale 
density and morphology.
Figure 3 shows that the early-type fraction $f_E$ 
monotonically increases as $\rho_{20}$ increases
for both early and late-type neighbor cases when $r_p \la r_{\rm vir,nei}$.
The fact that the background density dependence is manifest only when
a galaxy is sitting inside its neighbor's virial radius, rules out
the simple primordial origin scenario like that 
the relation is caused by the elliptical galaxies that preferentially
formed in higher density regions.
This is because Figure 3 shows isolated galaxies with $r_p > r_{\rm vir,nei}$
are almost ignorant of the background density.

One might also consider the effects of other neighbor galaxies, 
like the second and third nearest ones, which are monotonically increasing
as the background density increases.
But it again cannot explain why the virial radius of the nearest neighbor
should be the critical boundary for the onset of the background density
dependence.
Then, one would naturally suspect that some of the internal properties of 
the nearest neighbor galaxies depend on the background density 
in such a way that the strength of the neighbor influence is 
different at different background densities.

Park et al. (2008) proposed that the background density dependence
occurs due to the variation of hot halo gas property of galaxies.
In higher density regions, the halo gas of both early and late-type galaxies 
seems on average hotter and denser.
Galaxies with the same morphology, luminosity, and pair separation but, 
located in higher density regions are redder and less active than those 
in lower density regions (Fig. 7). This may be 
because the halo gas is maintained hotter and denser for the galaxies 
in higher density regions due to some internal heating mechanisms (like supernovae
and active galactic nuclei) 
and external confining material.
%A late-type galaxy with an early-type neighbor
%is redder and less active than a late-type galaxy 
%having the same luminosity, $r_p$, and even $\rho_{20}$
%but with a late-type neighbor. This may be because the halo gas 
%is relatively hotter and denser for an early-type neighbor 
%at given luminosity. 

This suggestion is yet to confirm observationally.
It is well-known that the X-ray luminosities $L_X$ of early types 
correlate with optical luminosity $L_B$ (see review by 
Mathews \& Brighenti 2003). More luminous 
early-type galaxies have larger $L{_X}/L{_B}$,
but with large scatter, than those of low luminosity.
The large scatter in the $L{_X}-L{_B}$ correlation has been
explained in terms of some specific environmental effect
on X-ray emission from galaxies
(Brighenti \& Mathew 1998; Brown \& Bregman 1998; 
Brown \& Bregman 2000; Helsdon et al. 2001; O'Sullivan et al. 2001;
Ellis \& O'Sullivan 2006; Jeltema et al. 2008).
Brown \& Bregman (2000) found that  
early types in denser environments have larger  $L{_X}-L{_B}$
and suggested that the X-ray luminosity is enhanced 
through accretion of the intergalactic gas or supression of galactic
winds by the ambient medium, whereas other authors found that 
galaxies in regions of high local galaxy density have 
lower $L{_X}/L_B$ (White \& Sarazin 1991; Henriksen \& Cousineau 1999).
Helsdon et al. (2001) and Matsushita (2001) suggested 
that the result of Brown \& Bregman is thought to be due to the large
fraction of luminous central-dominant group galaxies in the sample 
that are enhanced in $L_X$ due to an additional contribution from the 
intragroup or circumgalactic hot gas (e.g. group cooling flows) 
and isolated and non-central group galaxies show no significant
correlation between $L{_X}/L_B$ and environment.
%This means that there is the correlation between the hot gas of galaxy 
%and the background density in a manner.
%the halo gas
%of galaxies is hotter and denser in higher density environment. 
%This might have happened because of the higher pressure of the ambient medium
%or slower expansion of space in higher density regions, which can result in
%slower escapes of energetic particles generated within galaxies.
%and can affect the morphology and SF activity of the host and neighbor galaxies
%more strongly. 
%Having found the correlation between the hot gas property and the background
%density, 
%One can now explain the dependence of $f_E$ on $\rho_{20}$ when
%$r_p \la r_{\rm vir,nei}$.

Let us now consider the second major finding that more isolated galaxies 
are relatively brighter at fixed background density. 
It has been interpreted by Park et al. (2008) as evidence for transformation 
of galaxy luminosity class through the merger process.
The transformation rate through mergers should be higher in higher background
density regions and the dependence of luminosity on the pair separation 
be stronger in high density regions
(but note that we are not resolving massive clusters where 
mergers between ordinary galaxies are expected to be difficult to happen).
At a given background density morphology and
luminosity transformations can occur through galaxy interactions and mergers.
And the background density will statistically control the speed of the coupled 
evolution of morphology and luminosity. 
Considering the definite dependence of morphology and luminosity on
the neighbor distance at the present epoch and high redshifts (Hwang \& Park 2008), 
one can draw a conclusion that
these processes are the key galaxy evolution mechanisms in addition to
those like cold gas accretion and internal passive evolution which 
happen without resorting to neighboring galaxies.
% (but see Noeske et al. 2007a and 2007b).

Why is the correlation between luminosity and pair separation strong
only when the pair separation is larger than the virial radius of the neighbor?
It can be explained if recently merged galaxies have $r_p$ larger 
than $r_{\rm vir,nei}$ from the new nearest neighbor (i.e. a pair of galaxies
having vanishing $r_p$ merge each other and 
jump to large $r_p > r_{\rm vir,nei}$ after merger).
This is possible because a pair of galaxies would merge more easily
if they are located outside the virial radius of another larger galaxy.
Park et al. (2008) found, in a search for evidence for this interpretation, 
that at fixed background density the isolated 
galaxies with $r_p > r_{\rm vir,nei}$ show the post-merger features 
more frequently than those with $r_p < r_{\rm vir,nei}$,
particularly in high density regions. Therefore, among the galaxies 
located at the same background density, more-isolated ones are 
more likely to be recent merger products than less-isolated ones
and are likely to be brighter.
This does not mean isolated galaxies in general have experienced 
recent merger events. Isolated galaxies are preferentially located 
in low density regions where the merger rate is low, are on average 
expected to be passively evolving with less environmental influence.
Those who want to analyze passively evolving galaxies must sample
isolated galaxies located in low density regions only.

Our third major finding is that, once morphology and luminosity are fixed,
the remaining properties of galaxies are quite insensitive to the background
density, particularly when $r_p > r_{\rm vir,nei}$.
One noticeable exception is color. The color of late-type galaxies 
shows a weak residual dependence on the background density above the mean 
density even if both morphology and luminosity are fixed and even when
they are isolated.
If galaxies maintain hotter and denser halo gas in higher density regions
as we propose, it is possible for the SF activity of galaxies to drop
for a sufficiently long time and for galaxy color to become redder 
relative to those in lower density regions. Enhancement of SF activity
by late-type neighbor galaxies occurs at smaller neighbor separations
as the background density increases (see the E-l and L-l cases of the
$W(H\alpha)$ parameter in Fig. 7).
This observation can be also understood by the background 
density-dependent halo gas properties or 
mass transfer efficiency.

An interesting question regarding galaxy color is which is more
fundamental physical property of galaxies between morphology and color.
One way to address this question is whether or not galaxy morphology
shows any dependence on environment beyond its correlation to color
(Ball, Loveday \& Brunner 2008; van den Bergh 2007).
To answer this question we selected two local density subsets 
containing galaxies with $-19.5>M_r>-20.0$ and $2.6<u-r<3.0$ 
located at the background densities $\rho_{20}<{\bar\rho}$ 
(low-density subset) or $\rho_{20}>20{\bar\rho}$ (high-density 
subset).  Since we fixed both luminosity and color, the stellar 
mass of galaxies is effectively fixed (Yang et al. 2007).
The fraction of early types in the low-density subset is found
to be $0.76\pm 0.04$, but that in the high-density subset is 
$0.90\pm 0.04$.  Therefore, the morphology-density relation 
becomes much weaker when we severely limit both luminosity and 
color. But there is still some residual dependence of 
galaxy morphology on the background density.
This demonstrates morphology contains independent information on the 
environmental dependence of galaxies that color does not have.

According to the tidal interaction theory the energy deposit in a galaxy
is inversely proportional to the square of the
velocity difference between the interacting pairs (Binney \& Tremaine 1987).
Several previous works have reported a detection of such inverse 
correlation between the strength of interaction effects and the velocity 
difference between pairs (Barton et al. 2000; Lambas et al. 2003; 
Nikolic et al. 2004; Alonso et al. 2006; Woods et al. 2006).
We also find results consistent with the tidal picture from our accurate
morphology subsamples as shown in Figure 8;
the L-l galaxies, having smaller velocity difference than the L-e galaxies
(see Fig. 2),
show more variation in $c_{\rm in}$ and $\sigma$
with the neighbor separation than the L-e galaxies.

To address this issue directly we divided the sample of the L-l galaxies
into three subsets according to the velocity difference with neighbors 
$\Delta v$, and
measured the equivalent width of the $H\alpha$ line as a 
measure of the SF activity, and the central velocity dispersion as a
measure of galaxy structure, as a function of the neighbor separation
for each subset. Figure 10 shows the median relations and 68\% ranges
for these parameters.
The solid line is for the subset with $\Delta v <70$ km s$^{-1}$,
the dashed line for 70 km s$^{-1} \le \Delta v < 120$ km s$^{-1}$,
and the dotted line for  120 km s$^{-1} \le \Delta v < 400$ km s$^{-1}$.
Due to the small size of each subset the errors are large, but one can
still clearly see the relation is more sensitive to the neighbor separation
for pairs with smaller velocity difference.

Our result can be compared with those of Lambas et al. (2003) and
Alonso et al. (2006) who reported that the onset of interaction-induced
SF is seen for $\Delta v \la 350$ km s$^{-1}$. Nikolic et al. (2004)
reported the onset even when $\Delta v <900$ km s$^{-1}$. 
We find no significant enhancement for pairs with $\Delta v \ge 120$ km s$^{-1}$, 
only a small enhancement for those with $70\le \Delta v <120$
km s$^{-1}$, and a significant enhancement for $\Delta v <70$ km s$^{-1}$.
Figure 2 implies that the average morphology of galaxies 
with small $\Delta v$ is more likely to be late type
and the fraction of early types will increase as $\Delta v$ increases
until it reaches the field fraction.
So the average SF activity can appear to be higher for pairs with
smaller $\Delta v$ not because of the interaction effects but because of
higher fraction of late-type galaxies.
Most of the previous works did not carefully distinguish among different
morphological types of galaxies in pairs.
Therefore, it is likely that the results of the previous works are
contaminated by the average morphology variation with $\Delta v$
in addition to the genuine effects of interaction.
%Once we fix the morphology of the target and neighbor galaxies, 
%we can inspect the pure effects of interaction on the SF activity 
%for samples of pairs with different $\Delta v$'s.  

The dependence on $\Delta v$ 
is less strong for $\sigma$, but still shows up at the smallest 
separation bin. However, it is not clear whether or not the increase of 
$\sigma$ is entirely due to the matter perturbation within galaxies or
due to the additional contribution by the mass flow from the neighbor.

To check if our findings are robust against our choice of neighbor selection 
parameters, we redid our analyses using a sample of galaxies that are constrained to
have neighbors brighter than themselves (i.e. the limiting magnitude difference 
$\Delta M_r=0$ instead of 0.5). It was found that our findings remain true for these
galaxy pairs. 
%In addition, we also found 
%that the fainter companion in a pair of galaxies is more affected 
%by interactions. All SF and structure parameters show more variations
%with $r_p$ for the fainter ones. However, in the case of luminosity 
%more variation with $r_p$ is found for the brighter ones. 
%This can be just a statistical effect.
%Suppose a pair of galaxies is drawn from a set of galaxies with a magnitude
%range of $\Delta M$. If another pair are drawn from the same set but with 
%the magnitude range stretched to $2\Delta M$, 
%there will be on average more magnitude difference between 
%the brighter galaxies than between the fainter ones.

\begin{figure}
\epsscale{1.}
\plotone{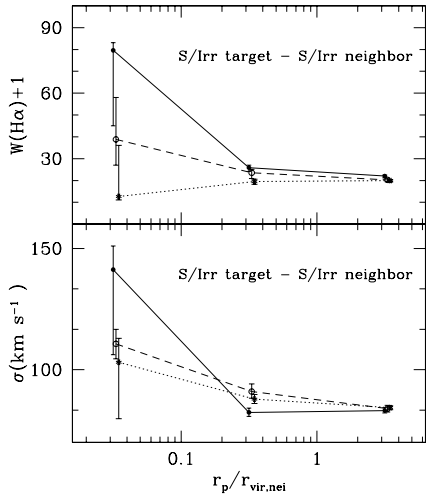}
\caption{The equivalent width of the $H\alpha$ line (top) and the central
velocity dispersion (bottom) of the L-l galaxies as a function of the neighbor
separation. The L-l galaxy subset is divided into three pairwise radial velocity
difference subsets; (solid) $\Delta v < 70$km s$^{-1}$,
(dashed) 70 km s$^{-1} \le \Delta v < 120$ km s$^{-1}$,
(dotted) 120 km s$^{-1} \le \Delta v < 400$ km s$^{-1}$.}
\end{figure}

\section{Conclusions}

We have inspected the dependence of eight physical parameters of galaxies
on the small-scale (the nearest neighbor distance, the nearest neighbor's 
morphology) and the large-scale (background density smoothed over 
20 nearest galaxies with $M_r < -19.0$) environments. 
We have also studied the kinematic properties of the galaxies in pairs.
We found the impact of interaction on galaxy properties are detectable at least
out to the pair separation corresponding to the virial radius
of (the neighbor) galaxies in our sample, which is mostly between
200 and 400 $h^{-1}$ kpc.
It was crucial to divide galaxy interactions into four cases depending on 
morphology of target and neighbor galaxies in order to detect these long-range
interaction effects.

Our major results are as follows.

1. There are two characteristic pair-separation scales where the
breaks in the dependence of galaxy properties on $r_p$ are observed.
The first scale is the virial radius of the nearest neighbor galaxy $r_{\rm vir,nei}$.
All parameters studied, except for $R_{\rm Pet}$, 
start to deviate from those of extremely isolated galaxies at 
$r_{p} \sim r_{\rm vir,nei}$ in the case of late-type galaxies, in particular. 
The second scale is the merger scale which is about $0.05 r_{\rm vir,nei}$. 
This corresponds to $10 - 20h^{-1}$ kpc for the galaxies in our sample.

2. The SF activity of galaxies is enhanced when the nearest neighbor is
a late type, but reduced when the neighbor is an early type.
These effects occur within the virial radius of the neighbor galaxy.
These are strong evidence for hydrodynamic interactions 
within the virial radius of the galaxy plus dark halo system during encounters.
%The HI gas is weakly bound by the galaxy's gravitational potential well, and
%can be easily removed (Boselli, \& Gavazzi 2006).

3. The dependence of galaxy properties on $\rho_{20}$ is strong
only for luminosity (Fig. 5) and morphology (Fig. 3).
All other parameters show weak or negligible dependence on $\rho_{20}$
once luminosity and morphology are fixed.
We have inspected the small subtle dependence on $\rho_{20}$ whenever detectable.
For example, the $u-r$ color of late-type galaxies has a weak residual dependence
on $\rho_{20}$ at fixed $r_p$.
We suggest that galaxies in higher density environment maintain
hotter and denser halo gas due to some internal heating mechanisms 
and external confining material, which can be the reason for the large-scale
density dependence of morphology and SF activity parameters.
 
4. At fixed large-scale background density, galaxies with larger pair 
separations have higher luminosity. 
Such dependence exists mainly at 
the nearest neighbor distance larger than the virial radius of the neighbor.
This is interpreted as evidence for the on-going process of luminosity 
transformation through mergers.

%As $r_{p}$ decreases below $r_{\rm vir,nei}$, the $u-r$ color and SF activity 
%slowly change in the direction determined by the morphology of the nearest neighbor 
%and the difference between the early- and late-type neighbor cases becomes larger.
%When $r_{p} \la 0.05 r_{\rm vir,nei}$, they start to diverge.
%The concentration and size of early-type galaxies have large dispersions, 
%and the concentration and central velocity dispersion of late-type galaxies 
%rise steeply at this merger scale. 

%For the parameters, $u-r$, $W(H\alpha)$, $\Delta(g-i)$, $c_{\rm in}$ and $\sigma$,
%we found that the impacts of interaction are larger for late-type galaxies.
%In particular, 
%5. The structural parameters $c_{\rm in}$, and $\sigma$, in particular, 
%showed the highest sensitivity to the neighbor separation when a late-type 
%galaxy encounters a late-type neighbor.
%Since the velocity difference between a galaxy and its neighbor is lowest for
%late-late pairs, according to the tidal theory, 
%they are expected to show the highest response to the interaction. 
%Therefore, our finding that the impacts of a galaxy-galaxy interaction depends 
%on the morphology of the target and neighbor galaxy, 
%This finding supports that the driving force of galaxy evolution 
%during galactic encounter is the tidal force.

In a forthcoming paper (Park \& Hwang 2008) 
we will examine the dependence of the SDSS galaxies associated with the
Abell clusters on the nearest neighbor separation and the clustercentric radius.
The latter is the large-scale environmental parameter replacing $\rho_{20}$ here.
This work will extend the present work to the extreme situation where 
the large-scale background density itself exceeds the virialized density.
We are also studying the effects of the nearest neighbor on galaxy properties
using higher redshift samples like the GOODS and DEEP2 samples (Hwang \&
Park 2008) to understand galaxy evolution due to galaxy-galaxy interactions 
in the high redshift universe.

\acknowledgments
The authors acknowledge the support of the Korea Science and Engineering
Foundation (KOSEF) through the Astrophysical Research Center for the
Structure and Evolution of the Cosmos (ARCSEC).
% and through the grant R01-2004-000-10520-0. 
%JRG is supported by NSF GRANT AST 04-06713.

Funding for the SDSS and SDSS-II has been provided by the Alfred P. Sloan 
Foundation, the Participating Institutions, the National Science 
Foundation, the U.S. Department of Energy, the National Aeronautics and 
Space Administration, the Japanese Monbukagakusho, the Max Planck 
Society, and the Higher Education Funding Council for England. 
The SDSS Web Site is http://www.sdss.org/.

The SDSS is managed by the Astrophysical Research Consortium for the 
Participating Institutions. The Participating Institutions are the 
American Museum of Natural History, Astrophysical Institute Potsdam, 
University of Basel, Cambridge University, Case Western Reserve University, 
University of Chicago, Drexel University, Fermilab, the Institute for 
Advanced Study, the Japan Participation Group, Johns Hopkins University, 
the Joint Institute for Nuclear Astrophysics, the Kavli Institute for 
Particle Astrophysics and Cosmology, the Korean Scientist Group, the 
Chinese Academy of Sciences (LAMOST), Los Alamos National Laboratory, 
the Max-Planck-Institute for Astronomy (MPIA), the Max-Planck-Institute 
for Astrophysics (MPA), New Mexico State University, Ohio State University, 
University of Pittsburgh, University of Portsmouth, Princeton University,
the United States Naval Observatory, and the University of Washington. 

{}
\end{document}